\documentclass[11pt]{article}
\usepackage[koi8-r]{inputenc}
\usepackage[russian]{babel}
\usepackage{epsfig}
\usepackage{amsmath}
\usepackage{amssymb}
\usepackage{longtable}
\newcommand{\ups}{\rule{0pt}{15pt}}
\def\mag{^m\!\!\!.\,}
\def\m{${}^m\!\!\!.\,$}
\def\asec{${}^{\prime\prime}$}

\title{ 
Optical Observations of Gamma-Ray Bursts, the Discovery
of Supernovae 2005bv, 2005ЕЕ, and 2006ak, and Searches
for Transients Using the "MASTER" Robotic Telescope
}

\author{ 
V.M. Lipunov, V.G. Kornilov,  A.V. Krylov, N.V. Tyurina,  A.A. Belinski, E.S. Gorbovskoy, D.A. Kuvshinov,   P.A. Gritsyk,  G.A. Antipov, G.V. Borisov, A.V. Sankovich,  V.V. Vladimiriov, V.I. Vybornov,  A.S. Kuznetsov
}

\begin{document}
\maketitle
\begin{abstract}
We present the results of observations obtained using theMASTER robotic
telescope in 2005.
2006, including the earliest observations of the optical emission of the
gamma-ray bursts GRB 050824 and
GRB 060926. Together with later observations, these data yield the
brightness-variation law $t^{-0.55+-0.05}$ for
GRB 050824. An optical flare was detected in GRB 060926.a brightness
enhancement that repeated the
behavior observed in the X-ray variations. The spectrum of GRB 060926 is
found to be $F_E \approx E^{-\beta}$, where
$\beta = 1.0 +- 0.2$. Limits on the optical brightnesses of 26 gamma-ray bursts
have been derived, 9 of these for
the first time. Data for more than 90
were taken and reduced in real
time during the survey. A database has been composed based on these data.
Limits have been placed on the
rate of optical flares that are not associated with detected gamma-ray
bursts, and on the opening angle for
the beams of gamma-ray bursts. Three new supernovae have been discovered: SN
2005bv (type Ia).the
first to be discovered on Russian territory, SN 2005ee --- one of the most
powerful type II supernovae known,
and SN 2006ak (type Ia). We have obtained an image of SN 2006X during the
growth stage and a light
curve that fully describes the brightness maximum and exponential decay. A
new method for searching for
optical transients of gamma-ray bursts detected using triangulation from
various spacecraft is proposed
and tested.
\end{abstract}

\newpage
{\renewcommand{\abstractname}{Abstract}{\relax}
\begin{abstract}
{\bf Optical observations of gamma-ray bursts, the discovery of supernovae 2005bv, 2005ee, 2006ak, search of transients at telescope-robot MASTER}

V.M. Lipunov, V.G. Kornilov,  A.V. Krylov, N.V. Tyurina,  A.A. Belinski, E.S. Gorbovskoy, D.A. Kuvshinov,   P.A. Gritsyk,  G.A. Antipov, G.V. Borisov, A.V. Sankovich,  V.V. Vladimiriov, V.I. Vybornov,  A.S. Kuznetsov

Sternberg astronomical institute, Moscow, Russia
Moscow state university, Moscow, Russia
Moscow Union ``Optic'', Moscow, Russia

The results of observations over 2005--2006 years at the robotic telescope MASTER are presented. There are the first in the world observation of optical emission of GRB050824 and GRB060926 gamma-ray bursts. Our data combined with more later one gives the low of brightness drop $t^{-0.55\pm 0.05}$ for GRB050824. We discovered optical flare for GRB060926 around 500--700 sec. The power low spectral index ($F_E \sim E^{-\beta}$) is equal $\beta = 1.0\pm0.2$. In the course of sky survey we have images of more than 90\% possible sky. The virtual data-base and pipe-line was made. The limit to the orphan optical bursts rate is presented. We discovered 3 supernovae stars, they are the following: SN2005bv (Ia-type) is the first one, opened from Russian territory, SN2005ee is one of the most powerful among II-type supernovae, SN2006ak (Ia-type). New method of the OT search after IPN-triangulation gamma-observation is proposed and tested.
\end{abstract}}
\newpage

\subsubsection*{INTRODUCTION}
The construction of robotic telescopes, which not
only automatically acquire but also automatically
process images and choose observing strategies, is
a new and vigorously developing area in modern
astronomy

MASTER (Mobile Astronomy System of TElescope
Robots), the first robotic telescope in Russia,
began to be created through the efforts of scientists
at the Sternberg Astronomical Institute of Moscow
State University and the Moscow .Optika. Association
in 2002, and continues to be developed to the
present [1, 2]. In its current form, the system has
four parallel telescopes on an automated equatorial
mount, capable of slewing at a rate of up to $6^{\circ}/s$
(located near Domodedovo, Moscow region), and two
very-wide-field cameras with separate mounts and
domes, with one located on the Mountain Solar Station
of the Pulkovo Observatory (near Kislavodsk),
roughly 1500 km from the other (which is also near
Domodedovo). The parameters of the telescopes and
CCD arrays are given in
~\ref{tabl:1}.

The system whose characteristics are closest to
the MASTER system
({\it http://observ.pereplet.ru})
 is the American ROTSE-III system [3] (http://
www.rotse.net). MASTER
(photo
~\ref{foto:1}) 
differs in its larger
field of view and the presence of several telescopes on
a single axis, which makes it possible to obtain images
at several different wavelengths simultaneously.
The main telescope, Telescope 1 (355 mm diameter,
a modified Richter-Slevogt system initially invented
by V. Yu. Terebizh) takes images in white light, and is
the main search element of the system. An Apogee
Alta U16 (
$4096 \times 4096$ 
pixels) is installed on this
telescope, making it possible to obtain images in a six
square degree field. A Sony video recorder is mounted
on Telescope 2 (200 mm diameter, a Richter-Slevogt
system constructed by G. V. Borisov), providing
images to 
$13 \div 14^m$
 with a time resolution of 0.05 s.
A prism is mounted at the focus of Telescope 3
(280 mm diameter, a Flugge system), providing
spectra of objects to $13^m$ in a $30^{\prime}\times 40^{\prime}$ 

Яfield of view
with a resolution of $50AA$  (with a Pictor-416 camera).
Telescope 4 (200 mm diameter, a Wright system
constructed by A. V. Sankovich) is equipped with
a filter cassette and SBIG ST-10XME camera. In
addition, MASTER has a very-wide-field camera
($50^{\circ}\times 60^{\circ}$), that covers the field of view of theHETE
orbiting gamma-ray telescope, making it possible
to obtain simultaneous observations with HETE to
$9^m$ using a separate automated scheme. This widefield
equipment enables searches for bright, transient
objects.

In the Summer of 2006, we installed a widefield
camera (MASTER-VWF-Kislovodsk) on the
Mountain Solar Station of the Main Astronomical
Observatory in Pulkovo, making it possible to continuously
monitor a 420-square-degree field of sky to
$13^m$ in a five second exposure.

Thus, we currently have three automated miniobservatories
with the following equipment:

\begin{itemize}\itemsep=-2pt
\item[--]Six telescopes and CCD arrays
\item[--]Three automated equatorial mounts.
\item[--]Three automated domes
\item[--]Two cloud-cover and temperature sensors
\item[--]One GPS receiver
\item[--]Six control and reduction computers
\end{itemize}

The Kislovodsk and Domodedovo systems are
connected via the Internet, and are able to respond
to the detection of uncataloged objects (optical
transients) within several tens of seconds (including
processing time). The results of observations using
the MASTER network will be reported separately.

MASTER is able to operate in a fully automated
regime: automatically, based on the ephemerides
(sunset) and the presence of satisfactory weather
conditions (the control computer is continuously
attached to a weather sensor), the roof (above the
main mount and wide-field camera) is opened, the
telescope is pointed at bright stars and pointing
corrections introduced, and, depending on the seeing,
it then either goes into a standby regime or begins a
survey of the sky using a specialized, fully automated
programme.

Thus, observations are conducted in two regimes:
survey and alert (e.g. observation of the locations)
of gamma-ray bursts based on coordinates obtained).
In the former case, the telescope automatically takes
three frames of an arbitrary region in succession, with
exposures from 30 to 60 s, moves to a neighboring
region 
$2^\circ$
away and carries out the same procedure,
and so on, repeating a given set of three frames every
40-50 min. This makes it possible to avoid artefacts
in the data processing, and to locate moving objects.
The alert regime is supported by a continuous connection
between the control computer and the GCN
international gamma-ray burst (GRB) network [4]
(http://gcn.gsfc.nasa.gov). After detection of a GRB
by a space gamma-ray observatory (SWIFT, HETE,
Konus-Wind, INTEGRAL etc.), the telescope obtains
the coordinates of the burst region (the socalled
coordinate error box), automatically points to
this direction, obtains an image of this region, reduces
this image, and identifies all objects not present in
the computer catalogs. If a GRB is detected during
the day, its coordinates are included in the observing
program for the next night.

A special program package for image reduction in
real time has been created, making it possible not only
to carry out astrometry and photometry of a frame,
but to recognize objects not contained in astronomical
catalogs: supernovae, new asteroids, optical transients,
and so forth.

Over the entire time observations have been obtained
on the MASTER system (see the results for
2002-2004 in [1,2]), images have been obtained for
52 GRB error boxes. In 23 cases, these observations
were the first in the world. In three cases, optical
emission was detected (this was the earliest detection
in Europe for GRB 030329, and the earliest in the
world for the two cases reported here).

Note that, from February through August 2006,
the main matrix of the search telescope was being
repaired, so that the sky survey essentially was not
conducted during this time.

\subsubsection*{OBSERVATIONS OF GRBS IN 2005-2006}
Starting at the beginning of 2005 through October
2006, the Domodedovo MASTER station carried out
observations of 31 GRBs
(see.~\ref{tabl:2}).
 In 16 cases, we
obtained the first upper limits on the optical fluxes of
the GRBs, i.e., fluxes brighter than which no optical
candidate for the GRB was detected.

Note that, before 2005, only a few alerts for GRBs
in the Moscow night-time sky in regions of sky accessible
to MASTER were received from the SWIFT
orbiting observatory, which provided more than 90%
of all detected GRBs in 2005, with one of these occurring
during rainy weather. Nevertheless, we were
able to report the first optical detections in the world
for two of these GRBs.

Unless otherwise indicated, we present instrumental
magnitudes in white light. Our photometry
was carried out in an automated regime using objects
in the frame identified with stars in the USNO-A2.0
catalog [51] (there were usually about 2000 stars in
a frame) and the combined USNO-A2.0. $R$ and $B$ 
magnitudes:

\begin{equation}
m = 0.89R + 0.11B
\end{equation}

We chose this combination so that our instrumental
magnitude was close to the $R$ magnitudes
of minor planets, i.e., objects with solar spectra. As
our observations show, these magnitudes are in poor
agreement with the $R$ magnitudes of GRBs, due to
the enhanced red sensitivity of the Apogee Alta U16
array..

The processing of a frame begins immediately after
it is taken, and requires less than one minute. As a
result, the robotic system can attempt to find unidentified
objects within the error box and compose the
text of a GCN telegram with the indicated brightness
limit for an optical transient. In parallel, our full
frame with the error box and an enlarged error box
(usually
 $6\div 8$ 
with an image of the same region in
the Palomar Sky Survey are sent to a database over
the Internet, together with our frames obtained during
the previous survey observations. Thus, an on-duty
observer can visually inspect the region to search for
objects with low signal-to-noise ratios (two to three).

If no object is found in the individual frames, the
images are summed. On a good, Moonless night, the
sum of 10-15 images lowers the limiting magnitude
to $20^m$. The results of our observations are summarized
in.~\ref{tabl:2}.

\subsubsection*{GRB 050824: EARLIEST IMAGE}

Information about the coordinates of GRB 050824
arrived at the MASTER observatory on August 24,
2005 after some delay due to the processing of the signal
in the SWIFT data center [52]. The first image of
the region was obtained 110 s seconds after obtaining
the alert, i.e., 764 s after the SWIFT detection (trigger
151905), during a nearly fully Moon. The optical limit
reached was $17.8^m$. The observations were continued
[35-38].
(~\ref{tabl:3}), 
and, during the reduction of the
summed frames and preparation of a telegram on the
results of our observations, telegram GCN3865 [53]
arrived with the coordinates of an
$18^m$ in $R$. This object was
present in our earlier summed images.

The earliest available optical image of this object,
obtained on our MASTER system, can be found
at the address http://observ.pereplet.ru/images/
GRB050824/1.jpg

~\ref{fig:1} presents the upper imits and magnitudes
obtained in the first minutes of observations.
~\ref{fig:2} shows our data together with the R data of the MDM
Observatory [56]. The MDM $R$ observations obtained
from 5.6 to 12.6 h after the GRB are consistent with
a power-law decay in the flux (with index 
$-0.55\pm0.05$
) 0.05), and also with our data obtained 24 min and
47 min after the burst.

\subsubsection*{THE SHORT BURST GRB 051103≈A SOFT
GAMMA-RAY REPEATER IN THE GALAXY
M81?}

GRB 051103 may be the first soft gamma-ray
repeater (SGR) detected outside our Galaxy; the first
image of the error-box region was obtained on the
MASTER telescope.

The bright, short (0.17 s) burst GRB 051103 was
detected by Konus-Wind, as well as HETE-Fregate,
Mars Odyssey (GRS and HEND), RHESSI, and
SWIFT-BAT [27]. The MASTER telescope started
to observe the error-box region for GRB 051103 [25]
several minutes after receiving the alert telegram [27].
The first image was obtained at 19 : 55 : 47 UT on
November 5, 2005, 2 d, 10 hr, and 30 min after the
GRB. We obtained 36 images with a total exposure
time of 1080 s between 19 : 55 : 47 and 21 : 45 :
17 UT. No optical transient was found in the error
box to $18\m5$ (in the presence of a full Moon and light
haze).

Analysis of the frame showed the presence of four
bright galaxies near or in the large error box 
(see photo ~\ref{foto:2} and .~\ref{tabl:4}).

The most likely candidate host galaxy is
M81 [27], and the burst itself has been interpreted as
a SGR. The error box lies outside any spiral arms,
where strongly magnetized neutron stars (magnetars,
which are thought to be the sources of SGRs) would
be likely to form. However, the structure of M81 is
distorted by tidal interaction, and part of a disrupted
spiral may fall in the error box. For example, the
ultraluminous X-ray source (ULX) M81 X-9 [57] is
located at a similar distance from the center of M81
(on the side opposite to the error box), and belongs to
that galaxy▓s population of massive stars. It would be
interesting to search for supernova remnants within
the error box (unfortunately, the supernova-remnant
survey [58] does not encompass the error box of the
GRB).

In our telegram [25], we also noted the elliptical
galaxy PGC 028505, which is close to the center of
the triangulation error box. Its distance is estimated
to be 80 Mpc. If the GRB occurred in PGC 028505,
the isotropic energy of the burst can be estimated
to be
$\sim 2\cdot 10^{49}$
erg. This exceeds the energy of the
short burst GRB 050509b, which is associated with
an elliptical galaxy [59], by an order of magnitude, but
remains fairly characteristic of long GRB energies.
Based on our data, Holland et al. [60] carried out
photometry ofPGC 028505 the following night, without
finding any optical transient brighter than $21\m$
although this work excluded the region of the galactic
bulge.

Nevertheless, the absence of an optical object represents
an additional argument that GRB 051103 is
the first SGR observed beyond our Galaxy, in the galaxy M81 [61]. Our full image of the errorgox
region can be found at the address http://observ.
pereplet.ru/images/GRB051103.4/sum36.jpg.

\subsubsection*{GRB 060926: EARLIEST IMAGE
AND DETECTION OF AN OPTICAL BURST}

Our observations of GRB 060926, detected by
the SWIFT gamma-ray observatory [62], were carried
out in an automated regime under good weather
conditions [63]. The earliest image was obtained on
September 26, 2006 at 16 : 49 : 57 UT, 76 s after
the detection of the burst. We found an optical transient
in our first and subsequent summed frames,
at the position
$R.A. = 17^h35^m43^s\!\!.66\pm 0^s\!\!.05, Dec.  = 13^\circ02^{\prime}18^{\prime\prime}\!.3\pm 0^{\prime\prime}\!.7$
, which coincides within the errors
with the coordinates of the optical transient reported
in [62]. Our photometry of the object provided the
earliest points on the light curve
(table.~\ref{tabl:5})

Our preliminary reduction indicated a more gradual
brightness decrease than the OPTIMA-Burst observations
[64] (the power-law index for the brightness
decrease in the first 10 min was 0.69). However,
subjecting the data to finer time binning revealed an
optical burst: after its initial decrease, the brightness
began to increase beginning 300 s after the
GRB, reaching a maximum500√700 s after theGRB
 (ЯЛ. ПХЯ.~\ref{fig:3}a). Synchronous SWIFT-XRT measurements
of the X-ray flux show similar behavior
(see fig..~\ref{fig:3}b).

Such an event had already been observed in at least
two other cases: GRB 060218A ($z = 0.03$) 1000 s
after the GRB [66], andGRB 060729 ($z = 0.54$) 450 s
after the GRB [67, 68]. Note that the GRB considered
here has a redshift of 3.208 [69].

The absorption indicated by the X-ray data corresponds
to
$n_H = 2.2\cdot 10^{21}\mbox{ЯЛ}^{-2}$
of which 
$n_H = 7\cdot 10^{20}\mbox{ЯЛ}^{-2}$
 occurs in the Galaxy [65]. Taking into
account the redshift, the total absorption in our band
should be
$\approx 3^m$
. Naturally, we assume here that the
dust-to-hydrogen ratio is the same as it is in our
Galaxy.

Acomparison of our optical data with the SWIFTXRT
X-ray fluxes [65] can be used to determine
the slope
$\beta$
 of the electromagnetic spectrum (
$F_E \sim E^{-\beta}$
),which turned out to be constant within the
errors and equal to
$1.0\pm0.2$
[5], which coincides with
the corresponding value for the X-ray spectrum. The
earliest image of the optical transient can be found at
\it {http://observ.pereplet.ru/images/GRB060926/
GRB060926_1.jpg},
the sum of the five following frames at
\it {http://observ.pereplet.ru/images/GRB060926/
GRB060926_5.jpg},
and the sum of the ten following frames at
\it{http://observ.pereplet.ru/images/GRB060926/
GRB060926_10.jpg}. Images of this same region obtained during the
previous set of survey observations can be found at
\it {http://observ.pereplet.ru/images/ GRB060926/
GRB060926_2005.jpg}.

\subsubsection*{OPTICAL EMISSION OF GRBS
IN THE FIRST HOUR AFTER THE BURST.}
In this section, we consider the question of how
universal the behavior of the optical emission of
GRBs is with regard to their absolute luminosity
and time behavior. We collected observations of the
following GRBs with known redshifts in the first hour
after their onset (ПХЯ.~\ref{fig:4}).

\begin{tabular}{ll}
GRB990123 & ([70])\\
GRB021004 & ([71], [72])\\
GRB021211 & ([73-75])\\
GRB030723 & ([76])\\
GRB040924 & ([77-79])\\
GRB041006 & ([80], [81])\\
GRB050319 & ([83], [84])\\
GRB050401 & ([85], [86])\\
GRB050502 & ([87], [88])\\
GRB050525 & ([89-91])\\
GRB050730 & ([92-95])\\
GRB050820 & ([96-98])\\
GRB050824 & ([34,35,36,53])\\
GRB050904 & ([99], [100])\\
GRB050908 & ([101-106])\\
GRB050922C & ([107-110])\\
GRB051109 & ([111-116])\\
GRB051111 & ([117-122])\\
GRB060926 & ([5-7])\\
\end{tabular}

We then normalized the magnitudes to a single redshift
(to take into account their different distances; 
~\ref{fig:5})
  and gamma-ray flux (to take into account their
directional beaming, assuming the beaming for the
optical and gamma-ray emission is similar
.~\ref{fig:6}).

Since the largest number of afterglows have
power-law light curves, we chose as boundaries in the
synthetic light curves straight lines (in logarithmic
coordinates) of the form
\begin{equation}
f(t) = 2.5\cdot 0.8\cdot \log t + c = 2\cdot \log t  +c.
\end{equation}
where c = const. The width of the band bounding the
synthetic light curve is then
\begin{equation}
\Delta m = c_{max} - c_{min}
\end{equation}

It▓s obvious that, the smaller
$\Delta m$
the better the
model used to construct the synthetic light curve
describes the real physical situation.

We normalized the observed curves using the
mean weighted gamma-ray flux for the entire sample:
\begin{equation}
F_{\gamma_0} = \sum_{n_i}f_in_i/\sum_{n_i}n_i = 2.13\cdot 10^{-5} \mbox{ЩПЦ/ЯЛ}^2,
\end{equation}
where $f_i$ are the individual fluxes and, $n_i$ is the number
of GRBs with fluxes in the interval:
\begin{equation}
i\times 10^{-6} \le f_i \le (i+1)\times 10^{-6}
\end{equation}
The normalization in redshift ($z_0 = 1$) was carried out
using the formula\begin{equation}
m^{z_0}_{opt} = m^{z}_{opt} - 5\log \frac{d_l(z)}{d_l(z_0)} = m^{z}_{opt} - 5\log \frac{(1+z)I(z)}{(1+z_0)I(z_0)}, 
\end{equation}
where
\begin{equation}
I(z) = \int\limits^{(1+z)d^{-1/3}}_1\frac{dq}{\sqrt{q^3+1}}, \qquad d=0.3/0.7
\end{equation}

We obtained synthetic curves in the optical and
gamma-ray normalized to the same redshift and
gamma-ray flux using the formula:
\begin{equation}
m_{*} = m^z_\gamma - 2.5\log\frac{1+z}{1+z_0}+2.5\log\frac{F^z_\gamma}{F^{z_0}_{\gamma_0}}
\end{equation}
Correcting for interstellar absorption does not appreciably
affect the distribution.

Applying the above normalization substantial narrowed
the width of the synthetic light curve, although
$\Delta m$
remained fairly large. We assumed that the synchrotron
mechanism operates in GRB sources, suggesting
that the spectrum should have the form: 
$\lambda^\beta$.

Let us try to predict the value of 
$\beta$
based on the
tendency for the width of the synthetic curve normalized
in this way to narrow.We must take into account
the so-called $K$ correction:

\begin{equation}
K_z = 2.5\log(1+z) + 2.5\log \left(\int I(\lambda)s(\lambda)d\lambda/\int I(\frac{\lambda}{1+z})s(\lambda)d\lambda \right)
\end{equation}
Formula (8) then takes the form:
\begin{equation}
m_{*} = m^z_\gamma - 2.5\beta\log\frac{1+z}{1+z_0}+2.5\log\frac{F^z_\gamma}{F^{z_0}_{\gamma_0}}
\end{equation}

~\ref{fig:7}
shows the dependence of the width of the
synthetic curve on the spectral index. We can see a
weak minimum for
$\beta=-1$,
with the corresponding
value of
 $\Delta m$ 
equal to the width of the synthetic curve
when we do not include the $K$ correction. Consequently,
the $K$ corrections are too small to appreciably
influence the width of the curve.

Thus, by normalizing the light cuves using an
appropriate set of parameters, we have been able to
appreciably decrease the width of the synthetic light
curve. However, while the dependence of the optical
flux on the gamma-ray flux is as unique as was
thought initially, there must be some reason for the
width of the synthetic GRB light curve. In addition to
internal absorption in the host galaxy, another possible origin is the directional beaming of the GRB
jets, which most likely has different opening angles
in the optical and gamma-ray, and, thereby, different
intensity distributions as a function of the viewing
angle.

\subsubsection*{NEW METHOD FOR SEARCHING
FOR OPTICAL TRANSIENTS ASSOCIATED
WITH GRBS DETECTED
BY TRIANGULATION}

All attempts to detect optical emission from GRBs
before 1997 were unsuccessful for various reasons.
First, the main method for deriving coordinates was
triangulation using data from several different spacecraft.
As a rule, this method gave very large error
boxes (from several to hundreds of square degrees).
Second, even these coordinates became available, at
best, only several days after the GRB itself. With the
launch of gamma-ray observatories equipped with Xray
telescopes, the size of the error boxes and reduction
time decreased sharply, and interest in searching
for optical emission for triangulated GRBs fell.

Nevertheless, we suggest that the situation with
triangulated GRBs has now changed, due, first and
foremost, to the appearance of robotic search telescopes
that are able to inspect large regions of sky
in relatively short times. For example, there exist a
number of specialized projects aimed at searches for
supernovae, minor planets, comets, and dangerous
asteroids. Our idea is that these telescopes could
carry out specialized searches within large error boxes
provided by triangulation [14]. This is especially important
for three reasons. First, it has recently become
clear that at least long GRBs are associated
with supernovae. Second, optical emission has been
detected days after some short GRBs. Third, as a rule,
anomalous bright GRBs are not detected by SWIFT,
which is the main provider of information about the
coordinates ofGRBs with accuracies up to several arc
minutes.

In 2005 and 2006, in cooperation with the Konus-
Wind project, we carried out several surveys at an
early stage (from several hours to several days), when
the error boxes reached several tens of square degrees.
TheMASTER telescope is able to obtain frames of up
to 360 square degrees to
 $18^m$--$19^m$
in an hour. For reliable
transient searches, we usually use our standard
three-frame scheme, taking three frames in a six-square-degree field of view sequentially over several
minutes, and returning to the same area an hour or
more later. This appreciably enhances the depth and
reliability of the search and makes it possible to identify
asteroids, including previously unknown ones, but
decreases the survey speed to 60 square degrees per
hour.

The MASTER robotic telescope uses two automated
search algorithms: a search for optical transients
not associated with known galaxies, and a
search for supernovae, as unidentified objects in a
frame near known galaxies from the HyperLeda catalog
[123]. We will illustrate this method using our
search for an optical transient from GRB 060425 [14]
as an example.

The short, hard burst GRB 060425 was detected
on April 25, 2006 at 16 : 57 : 40 UT by SWIFTBAT,
Konus-Wind, Suzaku-WAM, RHESSI, and
INTEGRAL-SPI-ACS [124]. A telegram with a refined
error box 2.53 square degrees in size arrived on
April 26, 2006 at 23 : 50 : 41 UT. The Konus-Wind
group provided preliminary information on April 25
with a large error box having the formof a narrow area
with a length of $7^\circ$ (fig.~\ref{fig:8}).This enabled a search
of the error-box region to begin at 19 : 53 : 00 UT on
April 26, 2006, 1.12 days after the GRB at twilight.
We then repeated the error-box survey onApril 27 and
30 andMay 3, 2006. The results of these searches are
listed in(table.~\ref{tabl:6}).

Our reduction of these images provided an upper
limit of
$17\m0$
 for the optical emission from
GRB 060425 at the indicated times. In other words,
no new objects were found in the GRB 060425
error box, in either the field or near known galaxies.
However, in spite of this lack of success, we believe
that thismethod is capable of yielding fruit in the near
future.

\subsubsection*{SEARCH FOR OPTICAL TRANSIENTS
NOT ASSOCIATED WITH DETECTED GRBS
(ORPHANED GRBS).}

Searches for ⌠orphaned■ GRBs is one of the most
important tasks of modern astronomy. Bursts from
such objects could be associated with GRBs that are
⌠hurtling past the Earth■ or could be the result of
other, completely new astrophysical processes.

Up to October 2006, we obtained roughly 80 000
images covering more than 90
no fewer than six times (Fig.~\ref{fig:9}).
In the reduced
MASTER images (and also those obtained at other
observatories), we encounter objects that have disappeared
40√50 min later, in the next series of frames.
They are clearly not asteroids, comets, spacecraft, or
other moving objects, and are not associated with
artefacts.

We have created a special section for the detection
of short-lived optical objects in the MASTER
database. The character of the sky survey (two series
of three frames, separated by 40√50 min), enabled
starting in June 2005 a continuous search for optical
transients. With the aim of automating the process
of distinguishing such objects and conducting more
trustworthy analyses, we adopted information about
the optical afterglows of GRBs as the basis for the
expected behavior of such objects, which will enable
us to place limits on the rate at which they appear.

According to the data [125], about 50
are brighter than 18.5m for the first 30 min after the
burst onset. This suggests the following transientsearch
criterion: a transient should be reliably detected
in all frames in the first series of images (three
frames with exposures of 30√60 s), but not in any
frame in the second series. This strategy will avoid
classifying various types of artefacts (⌠hot■ pixels,
cosmic rays, etc.) as transients. The use of a continuously
updated database on asteroids and other minor
planets enables us to also exclude these objects from
consideration as transients.

Unfortunately, we have thus far not been able to
demonstrate a trustworthy detection of an orphan
GRB. However, even this negative result can be used
to place limits on the frequency of GRBs. Following
[126], we will assume that a flash down to $17\m5$
 would be visible for 30 min. Since the surveys occur
at different times with different seeing, the observations
cannot all be considered to have equal weights.
Therefore, we determined the limiting magnitude for
each frame. The essence of the algorithm for estimating
the limiting magnitude is to determine the magnitude
for which there arises an appreciable difference
in the number of stars in the frame and in the USNOB
catalog [127]. Objects that have the limiting magnitude
found in this way have signal-to-noise ratios
of seven and higher. To ensure trustworthiness in the
results, the lowest of the limits in the frames was
adopted outside a given area.

We can estimate theminimumsolid angle in which
we would expect not to observe a single flash brighter 
than 
$17\m5$ 
during a year using the formula

\begin{equation}
\omega = S\times 30min\times \sum_i e^{m_i-17.5}
\end{equation}
where $\omega$  is the GRB frequency in (sq. deg.)╥year, $S$
is the size of the field of view, and $m_i$ the limiting
magnitude in area $i$. The field of view of theMASTE R
telescope after cutting off strips near the edge to filter
out false objects at the frame edge is $S = 6$ square
degrees. The summation is carried out over all survey
areas.

Fig.~\ref{fig:10}a and \ref{fig:10}b illustrates the distribution of limiting
magnitudes
$m_i$
for our survey. The limit on the GRB
frequency derived in this way using our data from
the MASTER telescope is 1.21 (sq. deg.)╥year to
($1^\circ)^2\cdot\mbox{year}$
  to
$17\m5$
Together with analogous results published by
the ROTSE group in [126], the joint limit is 2.95 (sq.
deg.)╥year.

\subsubsection*{SEARCH FOR SUPERNOVAE}

In [128], the rate of cosmological supernovae was
calculated based on a population-synthesis analysis
(the ⌠Scenario Machine■) for the first time. It was
shown that the rates of formation and detection of
distant supernovae depend directly on the fraction
of baryonic matter in the Universe present in stars
and the contribution of dark energy to the density
of the Universe. For example, based on data already
currently available, the most probable value for the
dark-energy contribution can be estimated as
 $\Omega_\lambda = 0.7$.

It can be shown using the results of these calculations
that, on average, in a six-square-degree field of
view, the number of supernovae at any given time $N$
will be
\begin{equation}
N = 2\cdot 10^{3(m-20)/5},
\end{equation}
where $m$ is the limiting magnitude for a frame in that
part of the sky.

We emphasize that this estimate was obtained
assuming an isotropic distribution of galaxies, and so
is applicable for supernovae located at distances of
more than 100 Mpc, i.e., for supernovae with maximum
brightnesses fainter than
$15^m-16^m$.
 Note, for
example, that the apparent magnitudes of Type Ia
supernovae located at redshifts
 $z \sim 1$ 
will be of the
order of $20^m$ 
(without including absorption).This
shows that even a random survey of the sky using a
wide-field telescope such asMASTER can lead to the
discovery of a substantial number of supernovae.

This is especially important in connection with
the study of distant Type Ia supernovae carried out
in [129]. It was shown that the distances to distant
galaxies determined from the estimated maximum
brightnesses of Type Ia supernovae occurring in them
were larger than the distances obtained from the redshifts
of lines in their spectra (the Hubble law). This
can be interpreted as evidence for the presence of a
vacuum energy (dark energy) that leads to an acceleration
of the expansion of the Universe. It is striking
that the contribution of the vacuum energy increases
with time, becoming substantial at redshifts
 $z<1$, 
so
that any observations of supernovae with magnitudes
brighter than
$20^m$, are important for verifying this
important discovery.

This last circumstance gave birth to a boom
in supernova searches. Currently, several hundred
supernovae are discovered at various observatories
around the world each year. However, as a rule, these
searches are conducted using telescopes with small
fields of view (less than one square degree) and lists
of galaxies that lead to certain selection effects. For
example, as a rule, the observations concentrate on
bright, giant galaxies,while themuchmore numerous
dwarf galaxies are largely omitted from the search.
Therefore, it is of considerable interest to carry out
a random search for supernovae using a wide-field
system such as MASTER.

We wrote a specialized program package for
the reduction of images to search for supernovae.
Searching for unidentified objects in a frame containing
more than 10,000 stars that are observed against
a non-uniform Galactic background is a complex
algorithmic task, which we have been able to solve
successfully.

When searching for new objects during a sky survey
with the MASTER system, all known and unknown
objects are distinguished automatically using
the program package we have devised. Searching
for supernovae in an automated fashion essentially
translates to identifying new objects near known
galaxies. The situation is greatly complicated by the
incompleteness of existing catalogs and the presence
of false objects in these catalogs. The general scheme
for finding supernovae in images that have passed
through the preliminary data processing is as follows:
(1) distinguish the signals of stellar objects above the
Galactic background, (2) compare the coordinates
andmagnitudes of these objects with those for objects
in catalogs, (3) if this area of sky has already been
observed by the MASTER telescope, compare any
new objects with those distinguished in the previous
survey≈if there is no known object at the coordinates
of a new object, it is taken to be a possible supernova.
The most complex part of this search procedure is
distinguishing the object against the Galactic background
and distinguishing between regions of star
formation and of supernovae.

Our survey of clusters of galaxies proved to be
efficient: we detected many supernovae in our fields
that had been discovered several days earlier. Due to
the deterioration of seeing in the area surrounding
Moscow in recent years, our rate of discovery of supernovae
is not very high; nevertheless, theMASTER
telescope has already discovered three supernovae,
two of them of Type Ia. These are the first supernovae
discovered on Russian territory.

We will now consider the supernovae discovered
by theMASTER telescope in more detail.

{\bf 2005bv} This was the first supernova discovered
in Russia [130].~\ref{foto:4}  shows the discovery
frame for SN 2005bv. It was discovered on April 28,
2005 as a result of the first survey by the MASTER
telescope, carried out over clusters of galaxies.
SN 2005bv is a Type Ia supernova with the coordinates: 
$R.A.(2000) = 14^h24^m07^s\!\!.44, Dec.(2000)  = +26^\circ17^{\prime}50^{\prime\prime}\!.3$, ГБЕГДМЮЪ БЕКХВХМЮ Б ДЕМЭ НРЙПШРХЪ 16\m8 (ОНКНЯЮ, АКХГЙЮЪ Й $R$ ТХКЭРПС). The supernova▓smagnitude on the day
it was discovered was 16.8m (in a band close to R).
The supernova is located 11\asec\ East and 9\asec\ South of
the galaxyPGC 1770866. The distance to SN2005bv
was measured in [131] based on galactic HII regions;
its redshift corresponds to a recessional velocity of
10 400 km/s.

{\bf 2005ee} АШКЮ НРЙПШРЮ 26 ЮБЦСЯРЮ 2005Ц, ГБЕГДМЮЪ БЕКХВХМЮ Б ДЕМЭ НРЙПШРХЪ --- 16\m0 (АКХГЙН Й $R$). нМЮ НРМНЯХРЯЪ ЙН II-РХОС [130], Х БЯОШУМСКЮ Б ЦЮКЮЙРХЙЕ PGC 73054 [131]. еЕ ЙННПДХМЮРШ $R.A.(2000) = 23^h57^m55^s\!\!.83, Dec.(2000)  = +32^\circ38^{\prime}08^{\prime\prime}\!.9$, Р. Е. SN2005ee МЮУНДХРЯЪ Б 3\asec\ Й ГЮОЮДС Х Б 5\asec\ North of
the center of PGC 73054 (its redshift corresponds
to 9730 km/s). The first estimates of its absolute
magnitude indicated that SN 2005ee is among the
most powerful of Type II supernovae that have been
studied in detail. Our long series of observations
testifies to the anomalously powerful plateau of this
supernovae. More detailed data and our theoretical
interpretation will be published separately ~\ref{foto:4} ОПЕДЯРЮБКЕМ ЯМХЛНЙ Я НРЙПШРХЕЛ ЯБЕПУМНБНИ 2006ee. presents the discovery frame for SN 2005ee. The
measured magnitudes and light curve are shown in table.~\ref{tabl:7} and on fig.~\ref{fig:11}.

{\bf 2006ak}
This supernova was discovered on
August 26, 2005; its magnitude on the day it was
discovered was $16.0^m$ (in a band close to R). It is a
Type II supernova [132] that exploded in the galaxy
PGC 73054 [133]. Its coordinates are
$R.A.(2000) = 11^h09^m32^s\!\!.83, Dec.(2000)  = +28^\circ37^{\prime}50^{\prime\prime}\!.3$, 
North of the galaxy
PGC 083454 (the redshift corresponds to
11 150 km/s [135]). The MASTER telescope obtained
four images of SN 2006 SN 2006ak  (table.~\ref{tabl:8})

{\bf 2006X}
In addition to the three supernovae
listed above, we obtained an image of the very bright
supernova SN 2006X in the galaxy M100 during our
survey, before the publication of its discovery [136]. Its
magnitude at epoch 06.06162 in February 2006 was
$16\mag2\pm0\mag3$. Our data point was the second to be
obtained on the rising branch of the Type Ia supernova
SN 2006X.~\ref{fig:12}, presents the photometric light
curve, and table ~\ref{tabl:9}. On fig ~\ref{foto:6} presents an
image of SN 2006X near maximum brightness.

The discovery of supernovae using the MASTER
telescope enables us to estimate the efficiency of
random searches for supernovae using wide-field
systems. According to our estimates, a telescope
with a diameter of 40√50 cm and a field of view of
six square degrees located under favorable seeing
conditions (for example at a height of 2000 m near
Kislovodsk) should be able to discover up to 100
supernovae per year. The MASTER-IV system (four
telescopes of the same type with an overall field of
view of 24 square degrees) would be able to discover
up to 500 supernovae per year, making it possible
to determine the contribution of dark energy to the
total mass of the Universe after two to three years of
observations.

\subsubsection*{CONCLUSION}
We have presented the results of observations obtained
on the MASTER robotic telescope in 2005√
2006, which is the only telescope of its kind in Russia.
These results include the first observations of optical
emission from the gamma-ray bursts GRB 050824
and GRB 060926. Our data together with observations
made later yield a brightness-variation law for
GRB 050824 of
 $t^{-0.55\pm 0.05}$. 
 During a survey, more
than 90
$19^m$. 
 is observed and
reduced, and a database created. We have been able
to place limits on the rate of optical flashes that are
not associated with GRBs, as well as on the beaming
angle for GRBs. We can place a limit on the rate
of detecton of GRBs to 17\m5, which corresponds
to 1.2 (sq. deg.) year. We have also discovered three
supernovae: SN 2005bv (Type Ia), which is the first
supernova discovered on Russian territory, the powerful
Type II supernova SN 2005ee, and SN 2006ak
(Type Ia).

Our experience of two years of operation of the
MASTER wide-field robotic telescope has demonstrated
its unique capabilities. If such systems could
be installed at suitable sites at various hour angles
across Russia, they would provide unique information
via continuous monitoring of both the near and distant
cosmos.

The authors thank the General Director of the
⌠OPTIKA■ Association S.M. Bodrov for providing
the MASTER project with necessary expensive
equipment. This work was partially supported by
the Russian Foundation for Basic Research (project
code 04-02-16411-Ю). The authors are grateful to
Dr. V.L. Afanas▓ev for useful discussions of the idea
behind the experiment and for kindly presenting us
with a prism, and also to the Konus-Wind group, and,
in particular, V. D. Pal▓shin, for collaborative work.
We also wish to thank the INET Internet provider
(
$http://inetcomm.ru/$
) for free access to the Internet
for the MASTER system in the Vostryakovo village
in Domodedovo, Moscow region.

\newpage

\begin{table}
\caption{Telescopes and receivers of the MASTER system in Domodedovo.\label{tabl:1}}
\centering
\bigskip
\begin{tabular}{rlrllrr}
\hline\hline
\ups &Telescope &Optical system &Diameter, mm & Light-gathering power& CCD array& Field\\
of view&Format,Mpixels\\[5pt]

\hline
\ups 1 & Richter√Slevogt& 355 & F/2.6 & Apogee U16E & $2.4^\circ\times2.4^\circ$ & 16\\
2 & Richter√Slevogt& 200 & F/2.6 & Sony LCL 902K & $1^\circ\times0.7^\circ$ & 0.4\\
3 &Flugge& 280 & F/2.5 & Pictor-416 & $1^\circ\times0.7^\circ$ & 0.4 \\
4 & Wright & 200 & F/4 & SBIG ST-10XME & $1^\circ\times0.7^\circ$  & 3.2 \\
5 & \parbox[t]{2.5cm}{Wide-field camera}& 25 & F/1.2 & \parbox[t]{4cm}{Foreman Electr. FE-285, Sony ICX285AL }&
$30^\circ\times40^\circ$ & 1.4 \\[18pt]
\hline\hline
\end{tabular}
\end{table}

\begin{center}
\begin{longtable}{l|l|l|l|l|l}

\caption{Observation GRB-error boxes in 2005-2006yy . \label{tabl:2}}\\
\hline\hline
\ups GRB & \parbox[t]{2.1cm}{Publication
in GCN
circular} & \parbox[t]{2.1cm}{Time from
detection of
the GRB} & \parbox[t]{1.4cm}{Limiting
magnitude} & \parbox[t]{1.3cm}{First
observation?} & \parbox[t]{5cm}{Comments}\\[32pt]
\hline
\ups GRB060926 & \parbox[t]{2.1cm}{5632 [5],\\ 5619 [6],\\ 5613 [7]} & 76~c & 17\m5 & 1 & \parbox[t]{5cm}{Optical transient detected, decay law
obtained.}\\

\ups GRB060712 & 5303 [8] & 212~Я & $14^m$ & 1 & \parbox[t]{4.5cm}{Pointing 72 s after the burst. No
optical candidate detected.}\\

\ups GRB060502B & 5056 [9] & 69~c, 82~m & $16^m$ &2 &\parbox[t]{5cm}{SWIFT GRB. Evening sky.Main
observations were 82 min later (after
sunset) with all MASTER
instruments: simultaneous $BVR$,
spectral, and high-time-resolution
images of the GRB region obtained.
No optical candidate detected.}\\

\ups GRB060427B & 5032 [10] & 18.5~h & & 1 & \parbox[t]{5cm}{Konus-wind GRB. Observations
began 18.5 hr after the GRB in the
survey regime (the delay was due to
processing of the signals from the
IPN triangulation equipment; the
IPN error box is a region tens of
square degrees in size). The region is
located near the Galactic plane.We
obtained roughly 20
six-square-degree images.}\\

\ups GRB060427 & 5020 [11] & 9~h 13~m & 17\m5 & 4 & \parbox[t]{5cm}{SWIFT GRB. First image obtained
on 04.27.2006 at 18:18:07 UT. No
new objects brighter than 17\m5 detected.}\\

\ups GRB060425 & \parbox[t]{2.1cm}{ 5026 [12],\\ 5008 [13],\\ 5080 [14]} & 26~h 53~min &
17\m5 & 1 & \parbox[t]{5cm}{ IPN triangulation GCN5005.
MASTER observed in the survey
regime on two nights: 04.26.2006
from 18:17:03 to 19:53:07 UT
(1.05√1.12 days after the event [13])
and 04.27.2006 from 18:57:51 to
19:21:20 UT. A total of 72 images.
were obtained. No new objects
brighter than15\m5 were detected.} \\

\ups GRB060421 & 4988 [15] & 562~s & 16\m8 & 3 & \parbox[t]{5cm}{Two minutes before the alert, the roof
was closed due to strong
cloud-cover. No new objects within
the SWIFT-XRT error box brighter
than 16\m8 $(S/N = 3)$ were detected.}\\

\ups GRB060319 & \parbox[t]{2.1cm}{4892 [16],\\ 4888 [17]}  & 162~s & 19\m5 & 2 &
\parbox[t]{5cm}{Spectral and integrated images of the
burst region were obtained. No new
objects brighter than 19\m5 were
found in the SWIFT error box.}\\

\ups GRB060213 & \parbox[t]{2.1cm}{4767 [18],\\4765 [19]} & 55~h 17~m & 17\m5 & 2 &
\parbox[t]{5cm}{GRB 060213 was detected by the
IPN. No optical candidate brighter
than 17\m5 was found.}\\

\ups GRB060211B &  4741 [20] & 282~s & 14\m5 & 2 & \parbox[t]{5cm}{Snow, gaps in the clouds.}\\

\ups GRB060209A &  4718 [21] & 235~s & 15\m8 & 2 & \parbox[t]{5cm}{No optical candidate found.}\\

\ups GRB060124 &  4572 [22] & 96~m & 14\m4 & 5 & \parbox[t]{5cm}{SWIFT GRB 060124. Information
about the GRB did not arrive through
the alert system. The weather was
hazy and cloudy.}\\

\ups GRB060118 & 4549 [23] & Synchronous & $8^m$ & 1 & \parbox[t]{5cm}{HETE alert 4006 without
coordinates. Observations before and
after the alert with the wide-field
camera. All exposures were 3 s. No
object brighter than $8^m$  detected.}\\

\ups GRB060111A & 4485 [24] & 130~s & $16^m$ & 4 & \parbox[t]{5.1cm}{Observations began half an hour
before dawn. No optical candidate
detected..}\\

\ups GRB051103 &  \parbox[t]{2.1cm}{4198 [25],\\4206 [26]} & 58~h 30~m & 18\m5 & 1 &
\parbox[t]{5.1cm}{IPN triangulation GRB 051103 [27].
Observations began several minutes
after receiving the alert. A total of 36
images with a total exposure of
1080 s were obtained. No optical
candidate brighter than 18.5m found
(the weather was hazy). Four
galaxies (M81,M82, PGC 2719634,
PGC 028505) are located near or in
the triangulation region; it is possible
the GRB arrived from a source in
PGC 028505.}\\

\ups GRB051028 & \parbox[t]{2.1cm}{4171 [28]\\4173 [29]\\4182 [30]} & 3~h 23~m & $17^m$--19\m4 & 2 & \parbox[t]{5.0cm}{HETE alert GRB 051028. No
objects brighter than 17\m9 (sum of
nine frames taken between 17:32:40
and 18:03:03 UT) and 19\m4 (8 h
26 m after the alert, total exposure of
1200 s) detected.}\\

\ups \parbox[t]{2.1cm}{Swift trigger 160640} & 4119 [31] & 1~В 11~ЛХМ & 16\m3 & 1 & \parbox[t]{5.0cm}{Technical alert. Sunset, haze. No
new object found based on
comparison with USNO-A2.}\\

\ups GRB051021.6 & 4118 [32] & 1~h 29~m & $14^m$ & 4 & \parbox[t]{5.0cm}{HETE alert 3947. Sunset, haze. No
new object found based on
comparison with USNO-A2..}\\

\ups GRB051011 & \parbox[t]{2.1cm}{4082 [33],\\4083 [34]} & 45~s & 17\m0 & 1 & 
\parbox[t]{5.0cm}{First image with exposure 5 s
obtained 45 s after detection of the
GRB. A secod image with exposure
30 s obtained 87 s after the GRB. No
new objects detected in the error box..}\\

\ups GRB050824 & \parbox[t]{2.1cm}{3886 [35],\\3883 [36],\\3869 [37],\\3870 [38]} & 764~s & 19\m4 & 1 & \parbox[t]{5.0cm}{Optical candidate detected.}\\

\ups GRB050825 & 3882 [39] & 94~s & 18\m9 & 1 & \parbox[t]{5.0cm}{SWIFT alert, no new objects found..}\\

\ups GRB050805b & 3769 [01] & 74~s & 17\m1 & 2 & \parbox[t]{5.0cm}{SWIFT trigger 149131. No new
objects found (MilkyWay region).}\\

\ups GRB050805a & 3767 [41] & 113~s & 14\m5 & 1 & \parbox[t]{5.0cm}{Weak object detected at the noise
level. Subsequently not confirmed..}\\

\ups GRB050803 & 3755 [42] & 198~s & 18\m6 & 2 & \parbox[t]{5.0cm}{SWIFT alert. No new objects found..}\\

\ups GRB050410 & 3221 [43] & 5~h & 18\m5 & 3 & \parbox[t]{5.0cm}{SWIFT alert. No new objects found..}\\

\ups GRB050408 & 3188 [44] & 1~h 09~m   & 14\m7 & 7 & \parbox[t]{5.0cm}{Sunset, cloudy; no new objects
found..}\\

\ups GRB050316 & 3108 [45] & \parbox[t]{2.0cm}{103~Я\\synchronous} & \parbox[t]{1.4cm}{
19\m5\\16\m5} & 1 & \parbox[t]{5.0cm}{No object was found on the summed
frame(19\m5) or the image obtained
from the survey observations
coincident with the GRB. This event
was later classified by the HETE
group as unconfirmed.}\\

\ups GRB050316 & 3106 [46] & 103~s & $18^m$ & 1 & \parbox[t]{5.0cm}{Preliminary result; 50 frames with an
exposure of 30 s. No new objected
detected based on a comparison with
USNO-B. A total of 50 spectra (50\AA) of a $40^\prime\times30^\prime$ region obtained.}\\

\ups GRB050126 & 2988 [47] & 2~h 48~m   & $17^m$ & 1 & \parbox[t]{5.0cm}{Object from GCN 2986 not detected..}\\

\ups GRB050126 & 2986 [48] & 2~h 48~m   & $15^m$ & 1 & \parbox[t]{5.0cm}{Optical transient at the noise level.
Not confirmed.}\\

\ups GRB050117 &  \parbox[t]{2.1cm}{2954 [49],\\2953 [50]} & 2~h & \parbox[t]{1.4cm}{$19^m$\\$17^m$} & 1 & \parbox[t]{5.0cm}{First observations of the region of
this SWIFT GRB..}\\[30pt]
\hline\hline
\end{longtable}
\end{center}

\begin{table}[h]
\caption{Observations of GRB 050824 .\label{tabl:3}}
\centering
\bigskip
\begin{tabular}{lllll}
\hline\hline
\ups Time (UT) & Time from GRB & Magnitude & Exposure & Comments\\[5pt]
\hline
\ups 23:25:00 & 788~s & >17\m8 & 45~s & Upper limit\\
23:25:00 -- 23:47:55 & 24~m   & $18\mag6\pm 0\mag3$ & $15\times30$~s & \\
23:49:00 -- 00:09:03 & 47~m   & $19\mag4\pm 0\mag3$ & $15\times30$~s & \\[4pt]
\hline\hline
\end{tabular}
\end{table}

\begin{table}[h]
\caption{Bright galaxies near and inside the large error box for GRB 051103 .\label{tabl:4}}
\centering
\bigskip
\begin{tabular}{lllllcc}
\hline\hline
\ups Name & Type & Coordinates (2000) & \multicolumn{2}{c}{\parbox[t]{2.7cm}{Magnitude\\
$B$~~~~{\small MASTER}}} & \parbox[t]{1.8cm}{Redshift} &  \parbox[t]{2cm}{Diameter of
$25^m$ isophote}\\[16pt]
\hline
M81 & Sab & $09^h55^m33^s\!\!.2$ +$69^\circ03^\prime55^{\prime\prime}$ & 7\m8 & --- & 0.000376 & large \ups \\
M82 & Sb  & $09^h55^m52^s\!\!.2$ +$69^\circ40^\prime47^{\prime\prime}$ & 9\m3 & --- & 0.000677 & large \\
PGC2719634 & --- & $09^h51^m32^s\!\!.3$ +$68^\circ31^\prime24^{\prime\prime}$ & 17\m8 & 16\m7 & --- & 16\asec9 \\
PGC028505 & E & $09^h53^m10^s\!\!.2$ +$69^\circ00^\prime02^{\prime\prime}$ & 17\m0 & 14\m8 & --- & 6\asec0 \\[4pt]
\hline\hline
\end{tabular}
\end{table}

\begin{table}[h]
\caption{Photometry of GRB 060926 obtained on the MASTER telescope.\label{tabl:5}}
\centering
\bigskip
\begin{tabular}{rrrrrr}
\hline\hline
\ups \parbox[t]{2.1cm}{Beginning of
exposure, s} & \parbox[t]{1.6cm}{Middle of
exposure, s} & \parbox[t]{1.8cm}{Exposure, s} & Magnitude & \parbox[t]{2.5cm}{Flux,
erg/($\mbox{sm}^2\cdot\mbox{Я}\cdot\mbox{eV}$)} & \parbox[t]{2.7cm}{Flux after correction for
absorption,erg/($\mbox{sm}^2\cdot\mbox{Я}\cdot\mbox{eV}$)} } \\[18pt]
\hline
\ups 76  &  91 & $1\times30$ & $17\mag3\pm 0\mag3$ & $(1.4\pm0.3)\cdot10^{-13}$ & $(4.3\pm1.0)\cdot10^{-12}$ \\
150 & 165 & $1\times30$ & $18\mag5\pm 0\mag3$ & $(4.6\pm1.1)\cdot10^{-14}$ & $(1.4\pm0.4)\cdot10^{-12}$ \\
165 & 343 & $5\times30$ & $19\mag3\pm0\mag3$ & $(2.2\pm0.5)\cdot10^{-14}$ & $(6.9\pm1.7)\cdot10^{-13}$\\
255 & 432 & $5\times30$ & $18\mag9\pm0\mag3$ & $(3.2\pm0.8)\cdot10^{-14}$ & $(9.9\pm2.4)\cdot10^{-13}$\\
343 & 519 & $5\times30$ & $18\mag5\pm0\mag3$ & $(4.6\pm1.1)\cdot10^{-14}$ & $(1.4\pm0.3)\cdot10^{-12}$\\
432 & 608 & $5\times30$ & $18\mag3\pm0\mag3$ & $(5.8\pm1.3)\cdot10^{-14}$ & $(1.7\pm0.4)\cdot10^{-12}$\\
519 & 707 & $5\times30$ & $18\mag4\pm0\mag3$ & $(5.1\pm1.2)\cdot10^{-14}$ & $(1.6\pm0.4)\cdot10^{-12}$\\
608 & 804 & $5\times30$ & $18\mag7\pm0\mag3$ & $(3.9\pm0.9)\cdot10^{-14}$ & $(1.2\pm0.3)\cdot10^{-12}$\\
707 & 1001& $5\times30$ & $20\mag0\pm0\mag3$ & $(1.2\pm0.3)\cdot10^{-14}$ & $(3.6\pm0.9)\cdot10^{-13}$\\
804 & 1200& $5\times30$ & $20\mag1\pm0\mag3$ & $(1.1\pm0.3)\cdot10^{-14}$ & $(3.3\pm0.8)\cdot10^{-13}$\\
901 & 1298& $5\times30$ & $>20\mag1\pm0\mag3$ &$<(1.1\pm0.3)\cdot10^{-14}$ & $<(3.3\pm0.8)\cdot10^{-13}$\\[4pt]
\hline\hline
\end{tabular}
\end{table}

\begin{table}[h]
\caption{Frames of the error-box region for GRB 060425.\label{tabl:6}}
\centering
\bigskip
\begin{tabular}{lccc}
\hline\hline
\ups Date & Time from GRB, days & Ref & Frame
limit \\[4pt]
\hline
\ups 26.04.2006 & 1.1 & [13] & 17\m5 \\
27.04.2006 & 2.1 & [12] & 17\m5 \\
30.04.2006 & 5.2 & ---  & 17\m8 \\
03.05.2006 & 8.1 & ---  & 17\m6 \\[4pt]
\hline\hline
\end{tabular}
\end{table}

\begin{table}[h]
\caption{Brightness of SN 2005ee.\label{tabl:7}}
\centering
\bigskip
\begin{tabular}{lcc}
\hline\hline
\ups Date (JD) & Magnitude & Uncertainty \\[4pt]
\hline
\ups 2453608.43819 & 16\m53 & 0\m04 \\
2453615.43489 & 16\m80 & 0\m03 \\
2453672.31788 & 17\m01 & 0\m06 \\
2453675.15890 & 16\m98 & 0\m02 \\
2453693.27430 & 17\m05 & 0\m02 \\
2453704.13819 & 17\m44 & 0\m07 \\[4pt]
\hline\hline
\end{tabular}
\end{table}

\begin{table}[h]
\caption{Brightness of SN 2006ak.\label{tabl:8}}
\centering
\bigskip
\begin{tabular}{lcccc}
\hline\hline
\ups Date (day in
February 2006) & Magnitude & Uncertainty & \parbox[t]{3.3cm}{Limiting magnitude in
frame } & Comments \\[16pt]
\hline
$08^d990$ & 17\m0 & 0\m3 & 17\m2 & \ups before discovery \\
$17^d970$ & 15\m8 & 0\m1 & 17\m5 & discovery \\
$17^d996$ & 15\m8 & 0\m1 & 17\m5 & \\
$17^d951$ & 15\m8 & 0\m1 & 17\m5 & \\[4pt]
\hline\hline
\end{tabular}
\end{table}

\begin{table}[h]
\caption{Brightness of SN 2006X.\label{tabl:9}}
\centering
\bigskip
\begin{tabular}{ll|ll|ll}
\hline\hline
\ups JD & Date (JD) & JD & Magnitude & JD & Magnitude \\[4pt]
\hline
\ups 2453772.5618 & 15.974 &  2453803.5534 &  14.173 &  2453849.3443 &  15.914\\
2453772.5626 & 15.960 &  2453803.5541 &  14.192 &  2453849.3450 &  15.861\\
2453775.4587 & 14.668 &  2453803.5549 &  14.134 &  2453849.3653 &  15.820\\
2453775.4595 & 14.714 &  2453812.4202 &  14.245 &  2453849.3660 &  15.916\\
2453775.4620 & 14.869 &  2453812.4212 &  14.271 &  2453849.3667 &  15.892\\
2453775.4632 & 14.676 &  2453812.4222 &  14.189 &  2453850.2783 &  15.869\\
2453775.4640 & 14.565 &  2453812.4901 &  14.205 &  2453850.2789 &  15.784\\
2453775.4648 & 14.921 &  2453813.2289 &  14.363 &  2453850.2794 &  16.073\\
2453775.4657 & 14.740 &  2453813.2307 &  14.361 &  2453850.2826 &  16.001\\
2453775.4666 & 14.894 &  2453813.2346 &  14.326 &  2453850.2833 &  15.781\\
2453775.4674 & 14.895 &  2453813.2405 &  14.382 &  2453850.2848 &  15.900\\
2453777.4437 & 14.411 &  2453815.3852 &  14.396 &  2453850.2987 &  15.947\\
2453777.4445 & 14.298 &  2453815.3870 &  14.383 &  2453850.2994 &  16.041\\
2453777.4618 & 14.311 &  2453815.3880 &  14.370 &  2453851.3209 &  15.983\\
2453777.4625 & 14.326 &  2453817.3577 &  14.520 &  2453851.3216 &  16.010\\
2453777.4633 & 14.315 &  2453817.3607 &  14.519 &  2453851.3425 &  15.954\\
2453777.4640 & 14.322 &  2453820.4075 &  14.789 &  2453851.3431 &  16.062\\
2453784.4029 & 13.602 &  2453820.4085 &  14.735 &  2453851.3438 &  15.931\\
2453784.4039 & 13.690 &  2453820.4094 &  14.741 &  2453853.3701 &  16.059\\
2453795.3588 & 13.896 &  2453820.4271 &  14.770 &  2453853.3708 &  15.986\\
2453795.3717 & 13.909 &  2453820.4281 &  14.761 &  2453853.3717 &  16.231\\
2453795.3726 & 13.905 &  2453820.4299 &  14.745 &  2453853.3962 &  15.987\\
2453795.3734 & 13.948 &  2453822.4769 &  14.862 &  2453853.3976 &  15.835\\
2453795.3762 & 13.927 &  2453822.4778 &  14.847 &  2453854.3684 &  16.004\\
2453798.4338 & 14.105 &  2453822.4799 &  14.885 &  2453854.3690 &  16.116\\
2453798.4364 & 14.120 &  2453831.4717 &  15.285 &  2453854.3902 &  15.985\\
2453798.4391 & 14.099 &  2453831.4726 &  15.287 &  2453855.4066 &  15.833\\
2453801.4549 & 14.184 &  2453831.4736 &  15.247 &  2453856.3672 &  16.063\\
2453801.4586 & 14.195 &  2453831.5024 &  15.251 &  2453856.3679 &  15.857\\
2453801.4604 & 14.158 &  2453845.3354 &  15.764 &  2453856.3686 &  16.082\\
2453801.5072 & 14.182 &  2453845.3362 &  15.756 &  2453856.3867 &  16.133\\
2453801.5080 & 14.186 &  2453845.3561 &  15.750 &  2453856.3874 &  16.032\\
2453801.5089 & 14.157 &  2453845.3568 &  15.832 &  2453856.3881 &  16.150\\
2453802.3225 & 14.222 &  2453845.3576 &  15.823 &  2453857.3728 &  16.338\\
2453802.4249 & 14.192 &  2453846.3363 &  15.647 &  2453857.3735 &  16.200\\
2453802.4289 & 14.169 &  2453846.3377 &  15.811 &  2453857.3988 &  16.054\\
2453803.4530 & 14.166 &  2453846.3537 &  15.716 &  2453857.3998 &  16.219\\
2453803.4540 & 14.126 &  2453848.3293 &  15.864 &  2453857.4009 &  16.276\\
2453803.5519 & 14.125 &  2453848.3300 &  15.841 &  2453860.3159 &  16.518\\
2453803.5526 & 14.230 &  2453848.3503 &  15.792 &  2453860.3169 &  16.349\\[4pt]
\hline\hline
\end{tabular}
\end{table}

%

\newpage
{\renewcommand{\figurename}{Photo}{\relax}
\begin{figure}[h]
\centerline{\psfig{figure=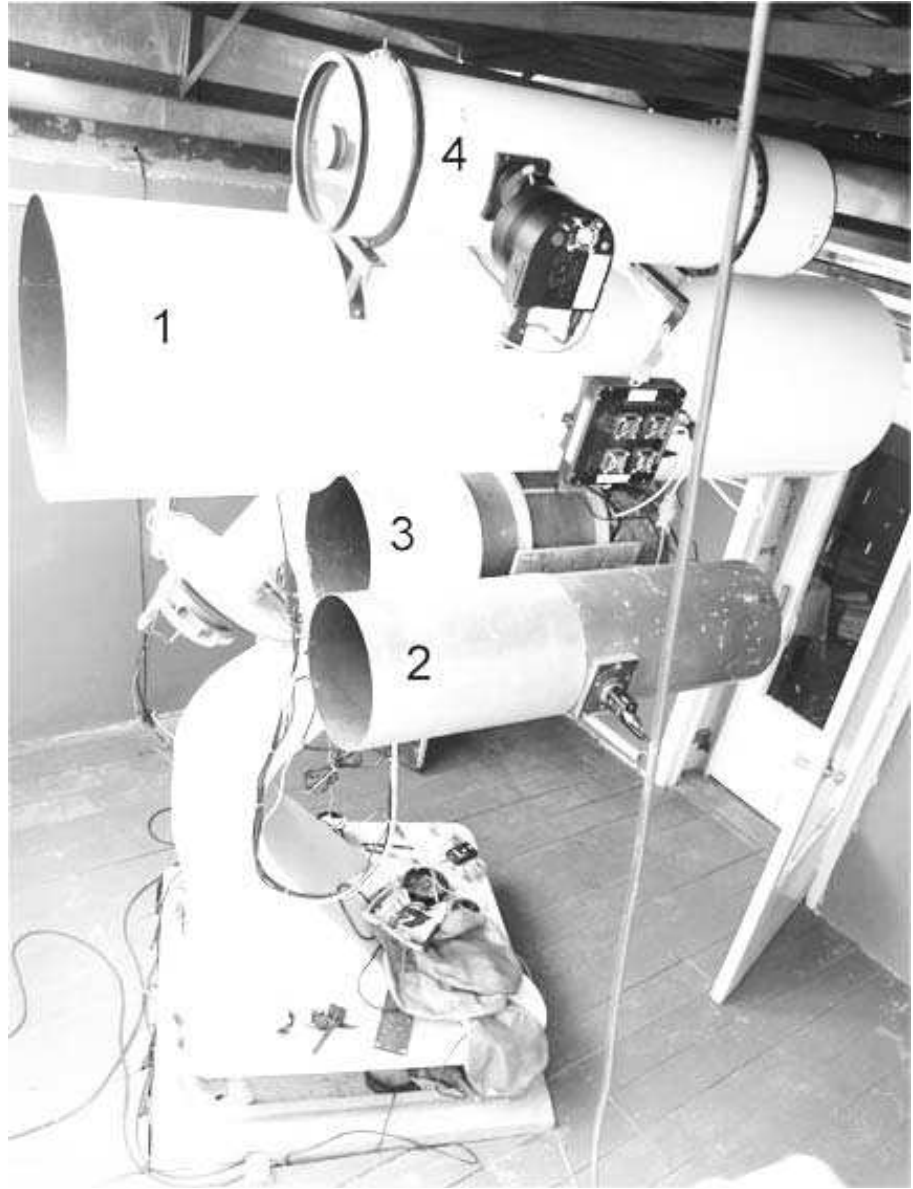,width=12cm}}
\caption{The MASTER wide-field robotic telescope in its observatory near Domodedovo (January 2006).\label{foto:1}}
\end{figure}

\begin{figure}[h]
\centerline{\psfig{figure=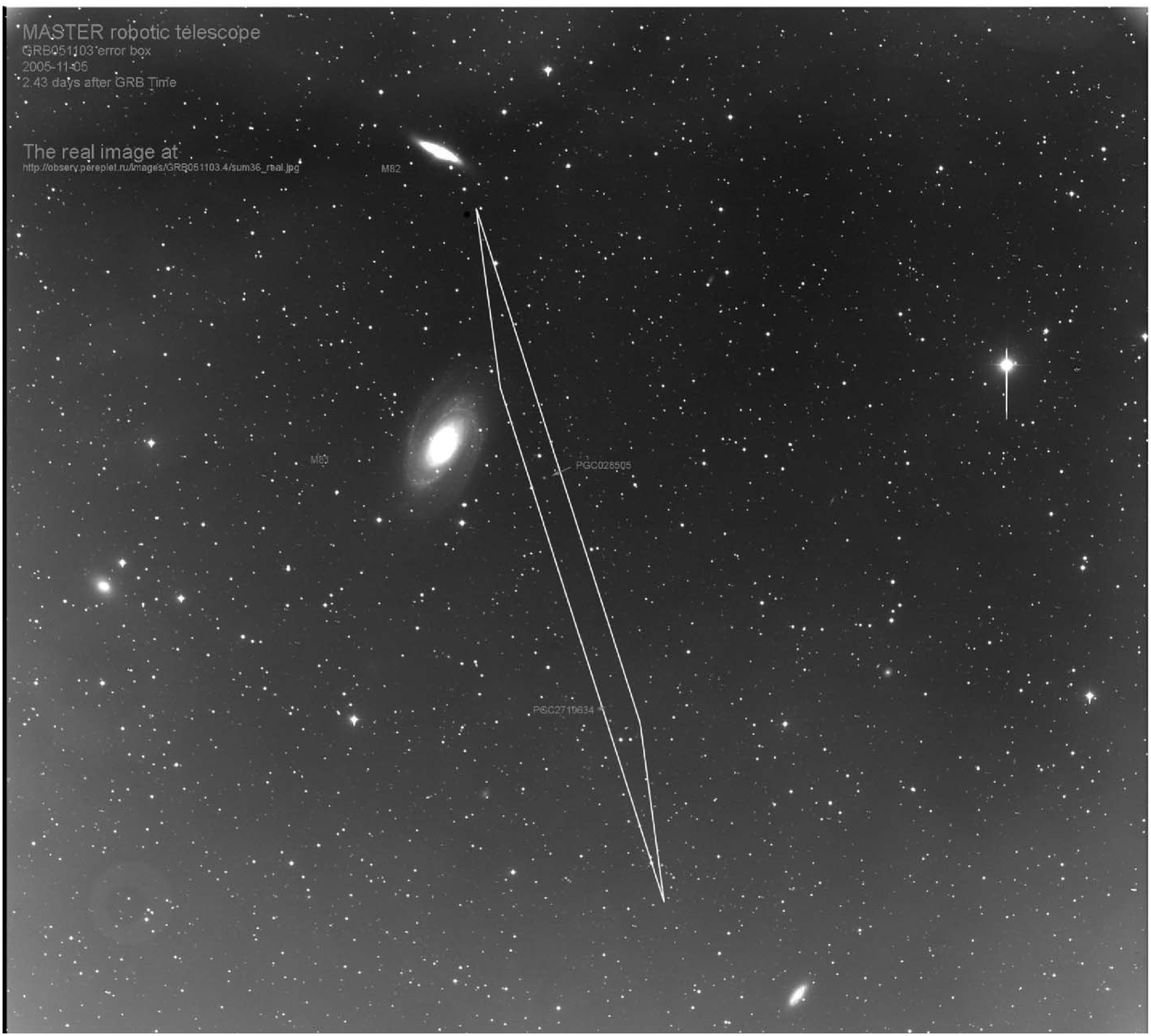,width=14cm}}
\caption{Image of the region of the short burst GRB 051103 (the error box is shown) obtained from a sum of 30
MASTER frames. The limiting optical magnitude in a frame is $18.5^m$. An image of the error box can be found at
http://observ.pereplet.ru/images/GRB051103.4/sum 36.jpg.} \label{foto:2}}
\end{figure}

\begin{figure}[h]
\centerline{\psfig{figure=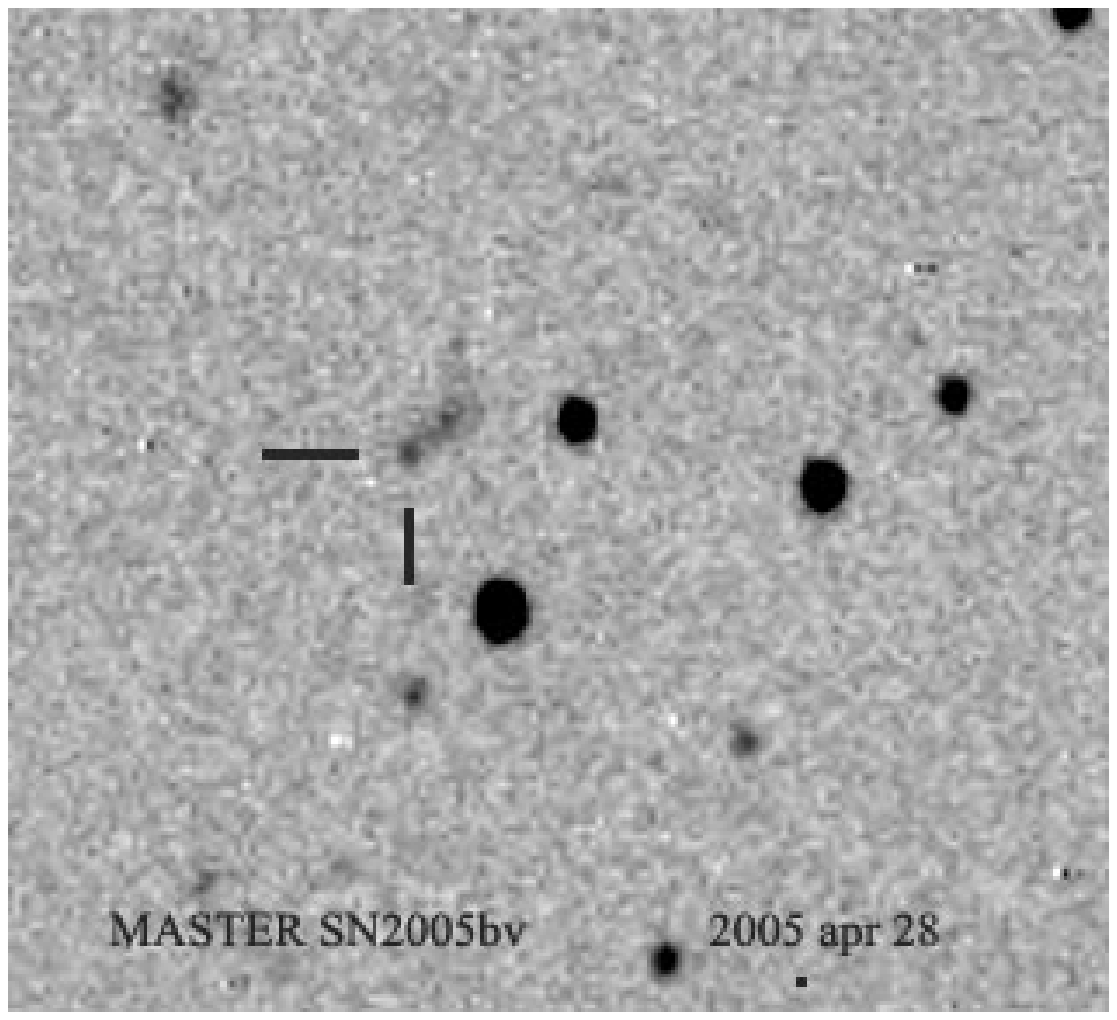,width=7cm}}
\caption{ Discovery frame for the supernova SN 2005bv \label{foto:3}}
\end{figure}

\begin{figure}[h]
\centerline{\psfig{figure=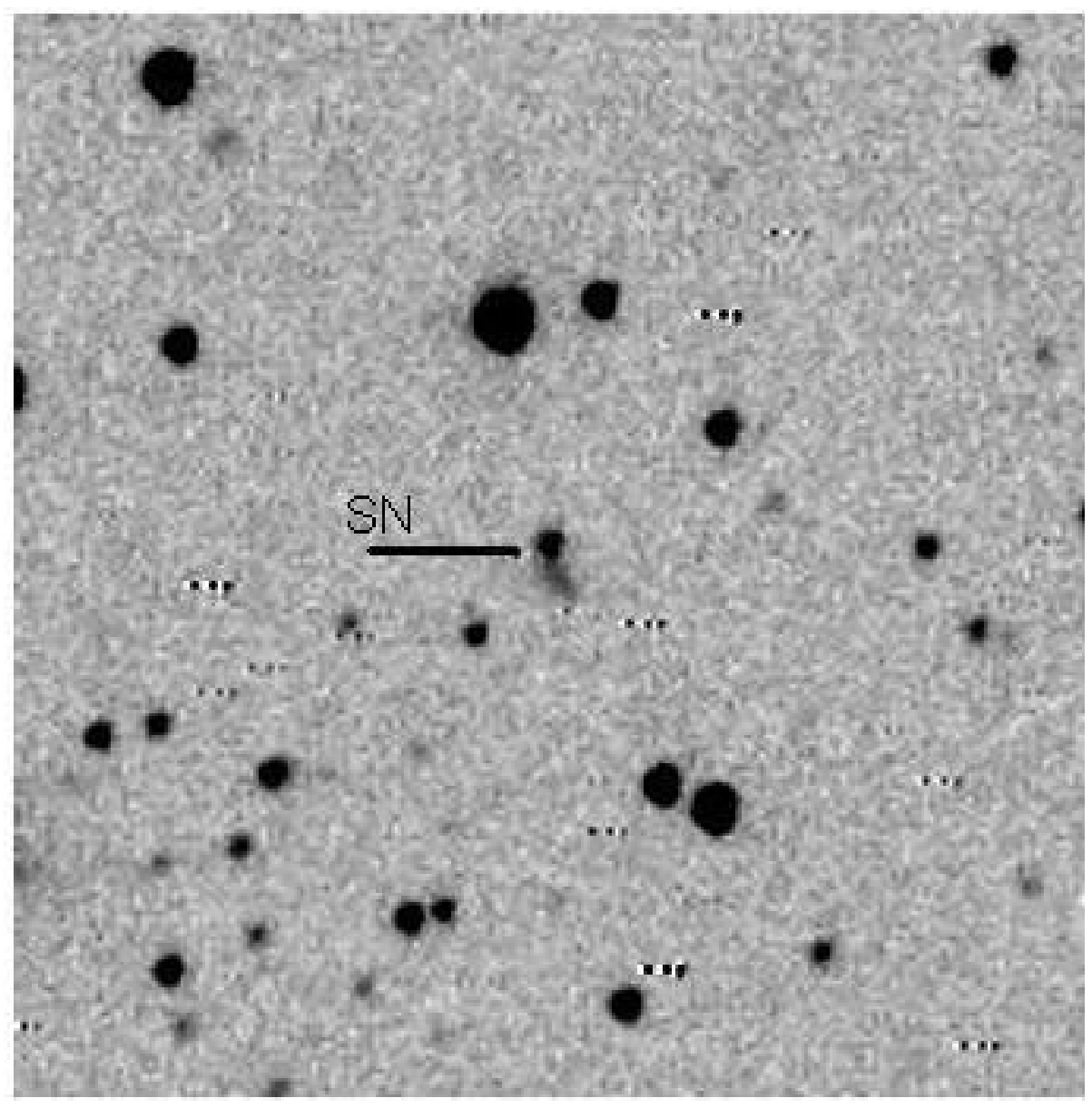,width=7cm}}
\caption{ Discovery frame for the supernova SN 2005ee \label{foto:4}}
\end{figure}

\begin{figure}[h]
\centerline{\psfig{figure=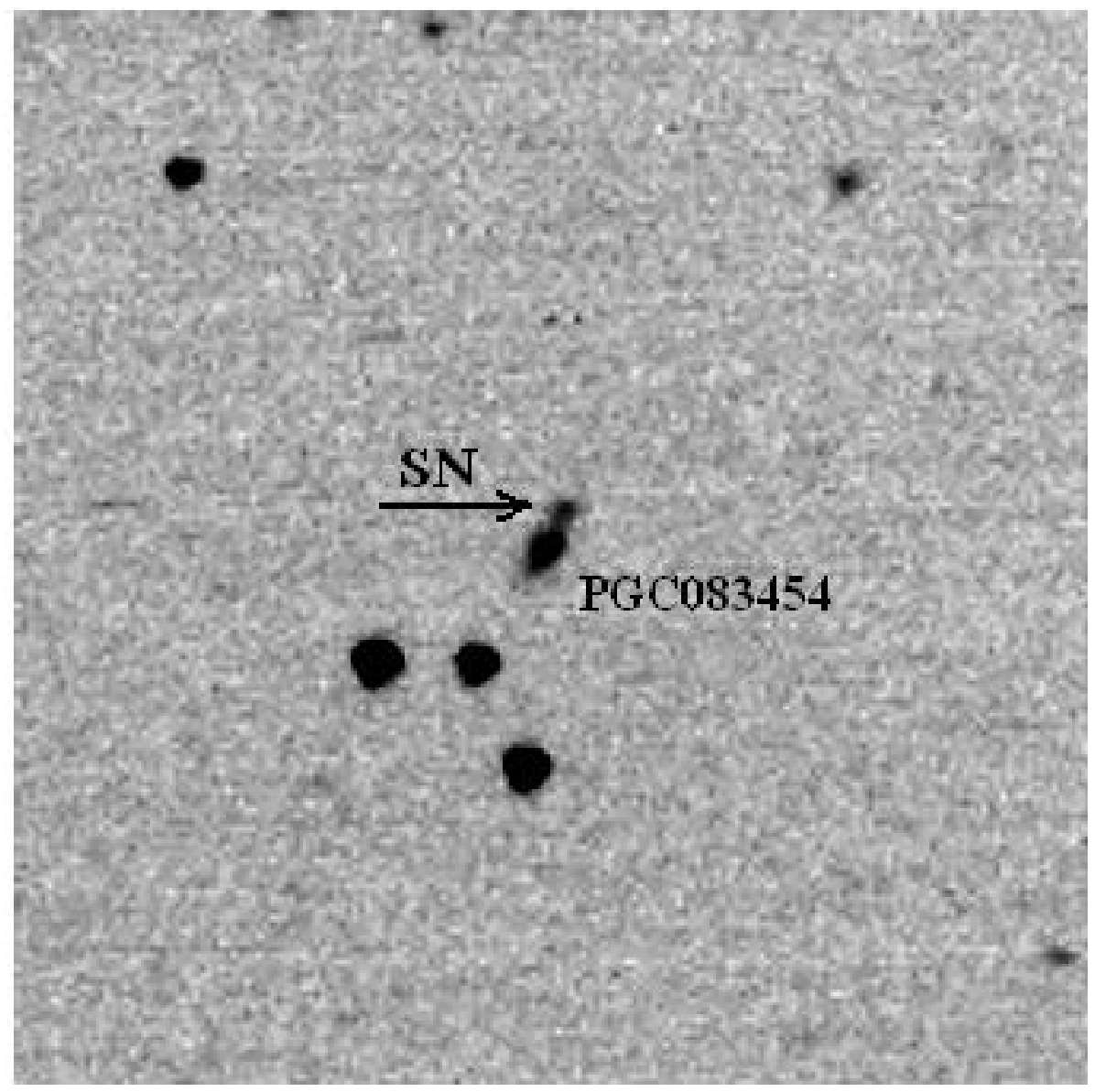,width=7cm}}
\caption{Discovery frame for the supernova SN 2006ak \label{foto:5}}
\end{figure}

\begin{figure}[h]
\centerline{\psfig{figure=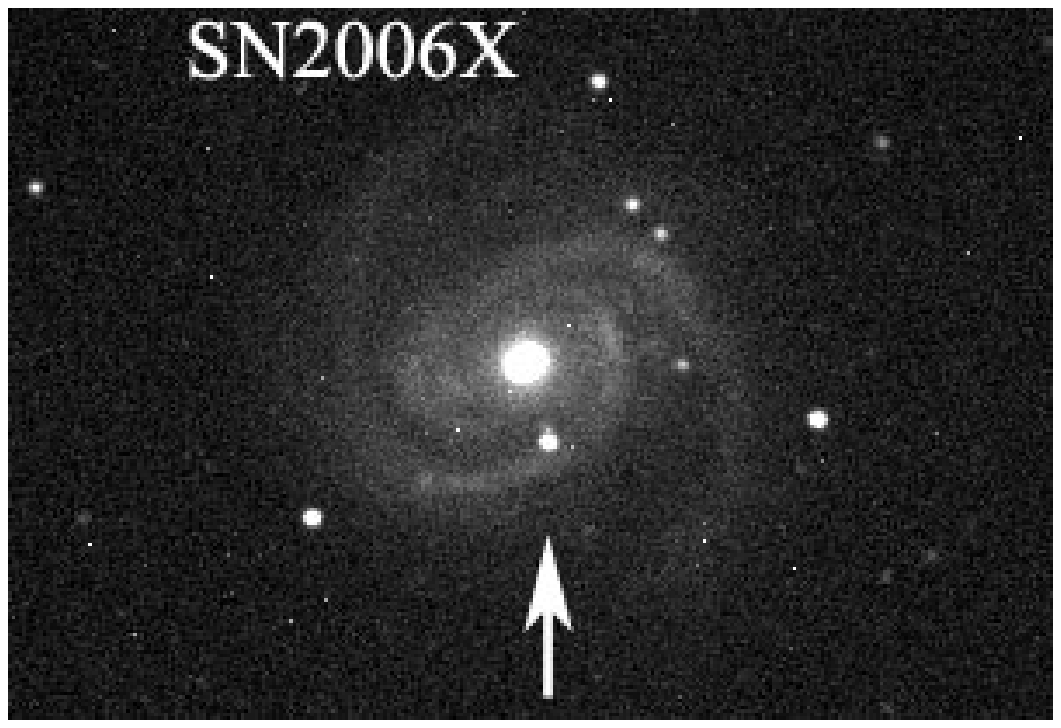,width=7cm}}
\caption{ Supernova SN 2006X (near maximum brightness)
in the galaxy M100 \label{foto:6}}
\end{figure}
}
\setcounter{figure}{0}

\begin{figure}[h]
\centerline{\psfig{figure=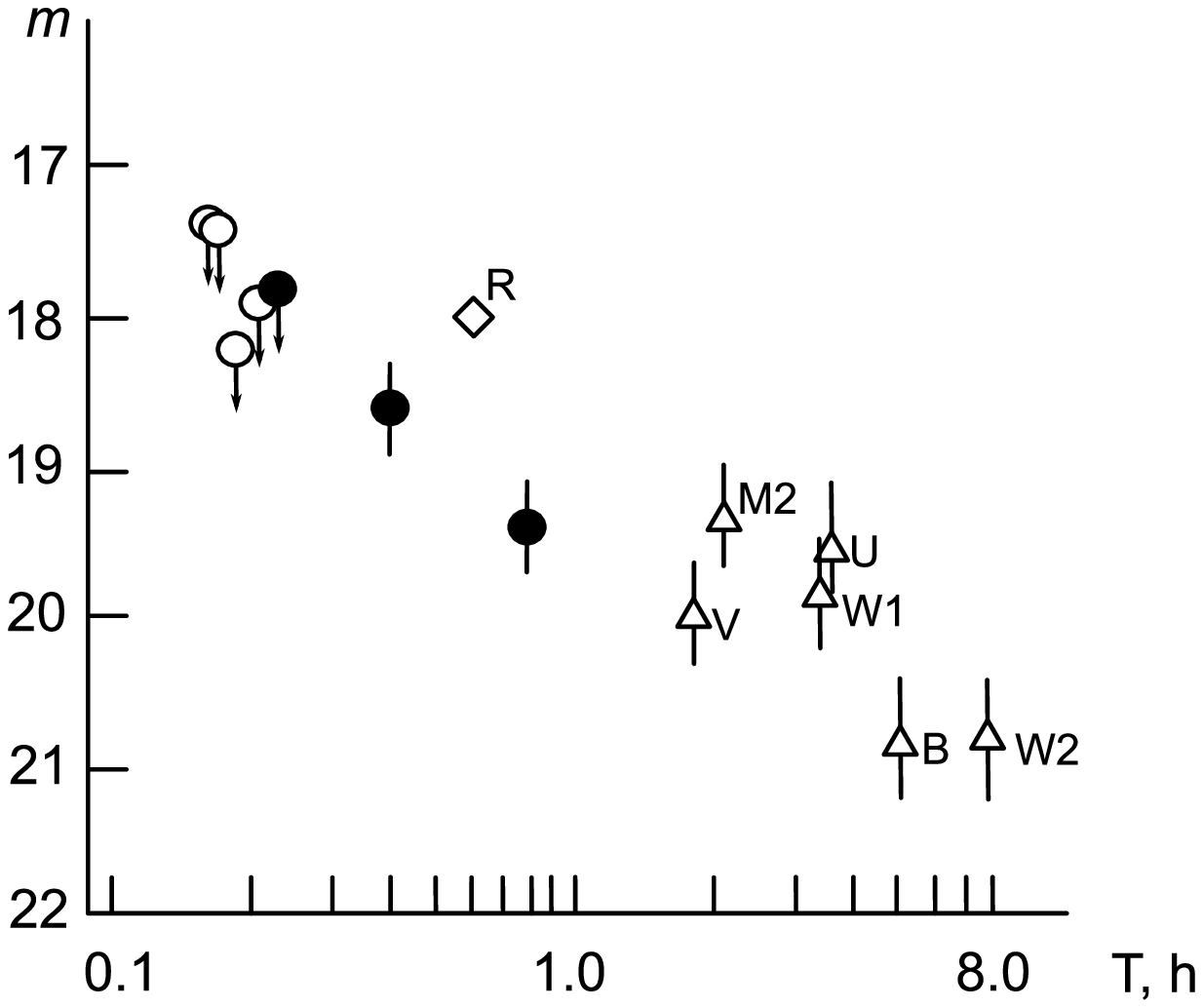,width=9cm}}
\caption{ Upper limit and brightness estimate for the optical
counterpart of GRB 050824 obtained in the first minutes
after the burst. The hollow circles are ROTSEestimates
[54], the dark circlesMASTER estimates [36], the
diamonds the $R$ magnitudes from [53], and the triangles
the SWIFT estimates in various photometric bands [55].\label{fig:1}}
\end{figure}

\begin{figure}[h]
\centerline{\psfig{figure=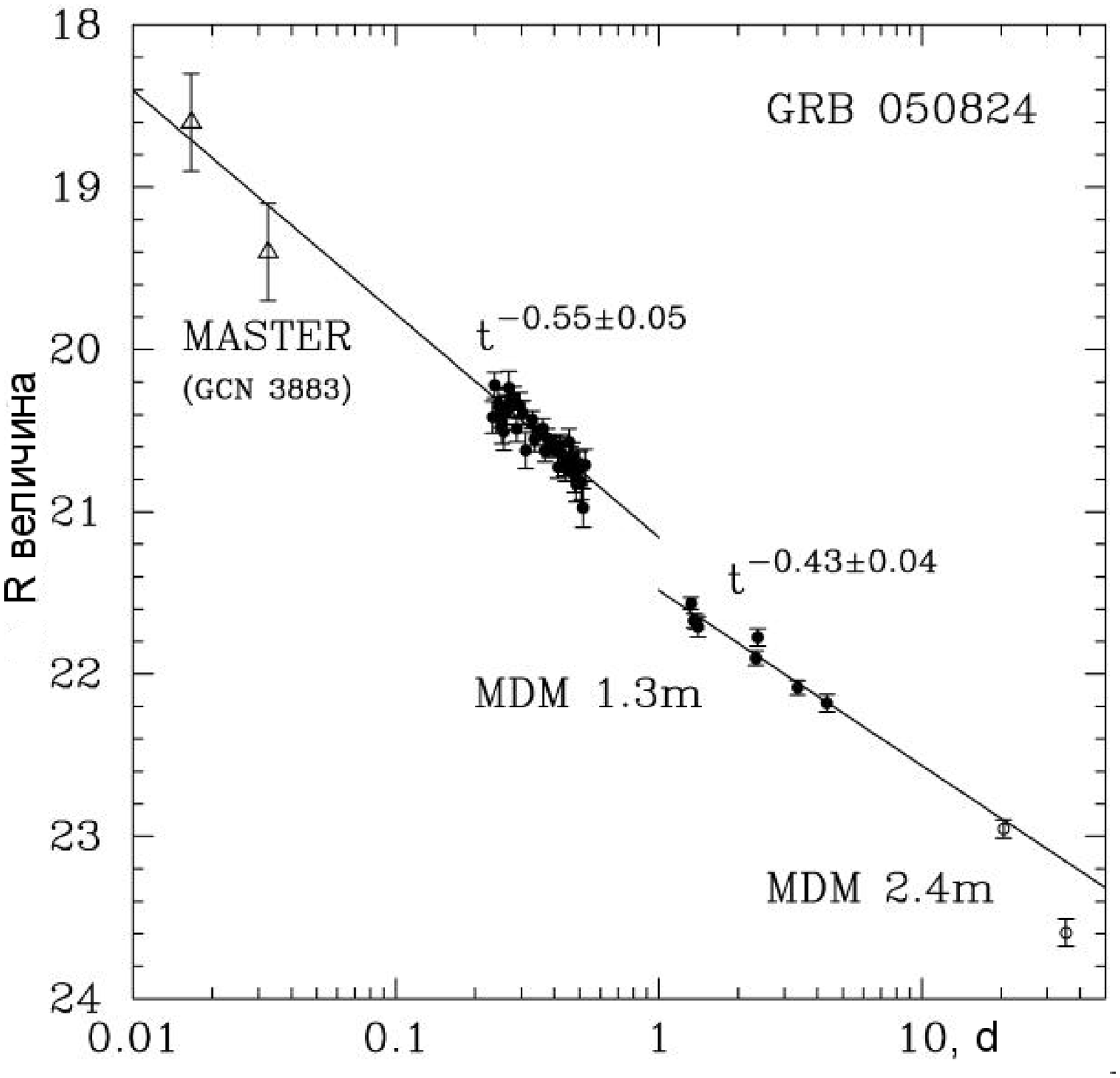,width=10cm}}
\caption{Observations of GRB 050824 from theMASTER telescope and the MDM Observatory ($R$) [56]. \label{fig:2}}
\end{figure}

\clearpage

\begin{figure}[h]
\centerline{\psfig{figure=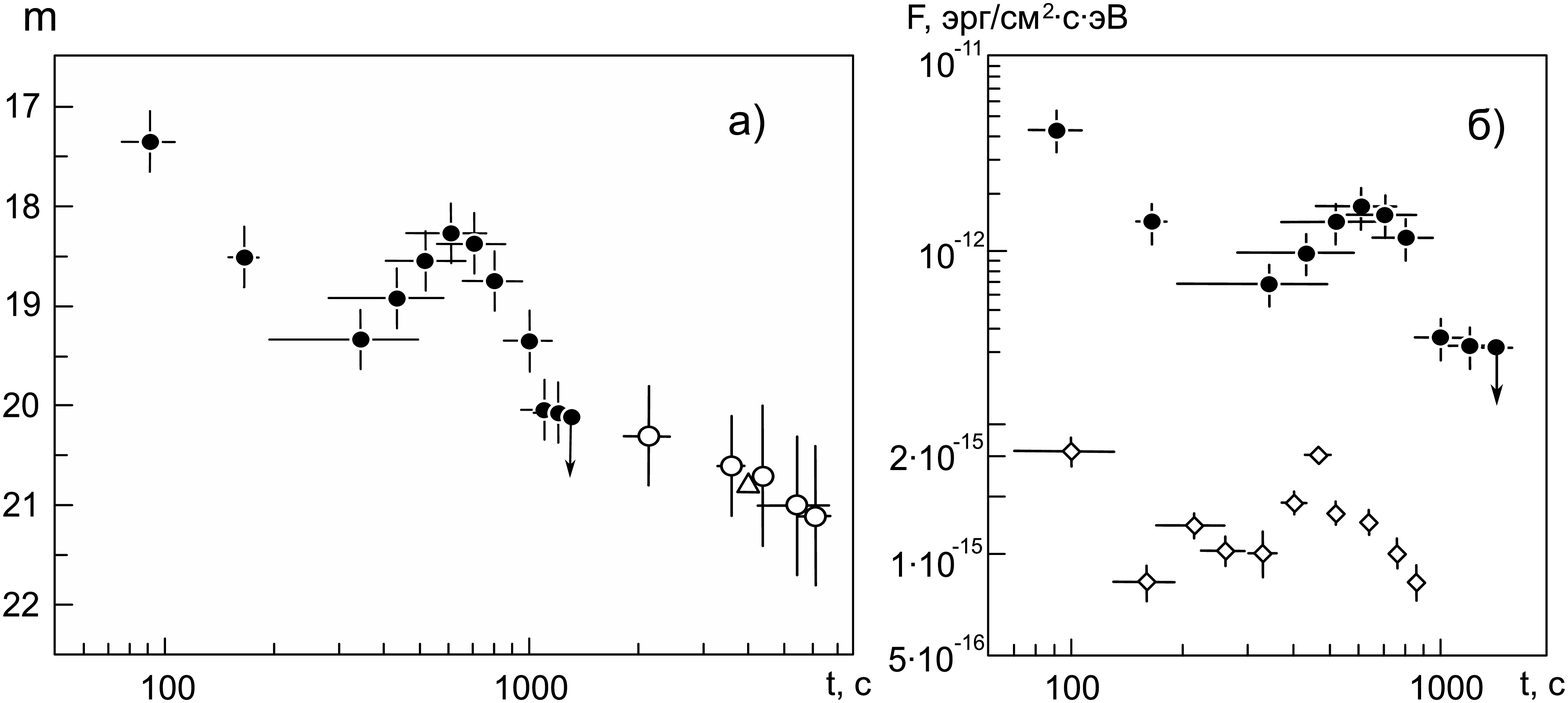,width=17cm}}
\caption{Comparison of the light curve of GRB 060926 obtained using the MASTER telescope [63] (points) with the (a)
OPTIMA Burst [64] (circles) and (b) SWIFT XRT (0.3√10 keV; diamonds) [65] light curves.\label{fig:3}}
\end{figure}

\begin{figure}[h]
\centerline{\psfig{figure=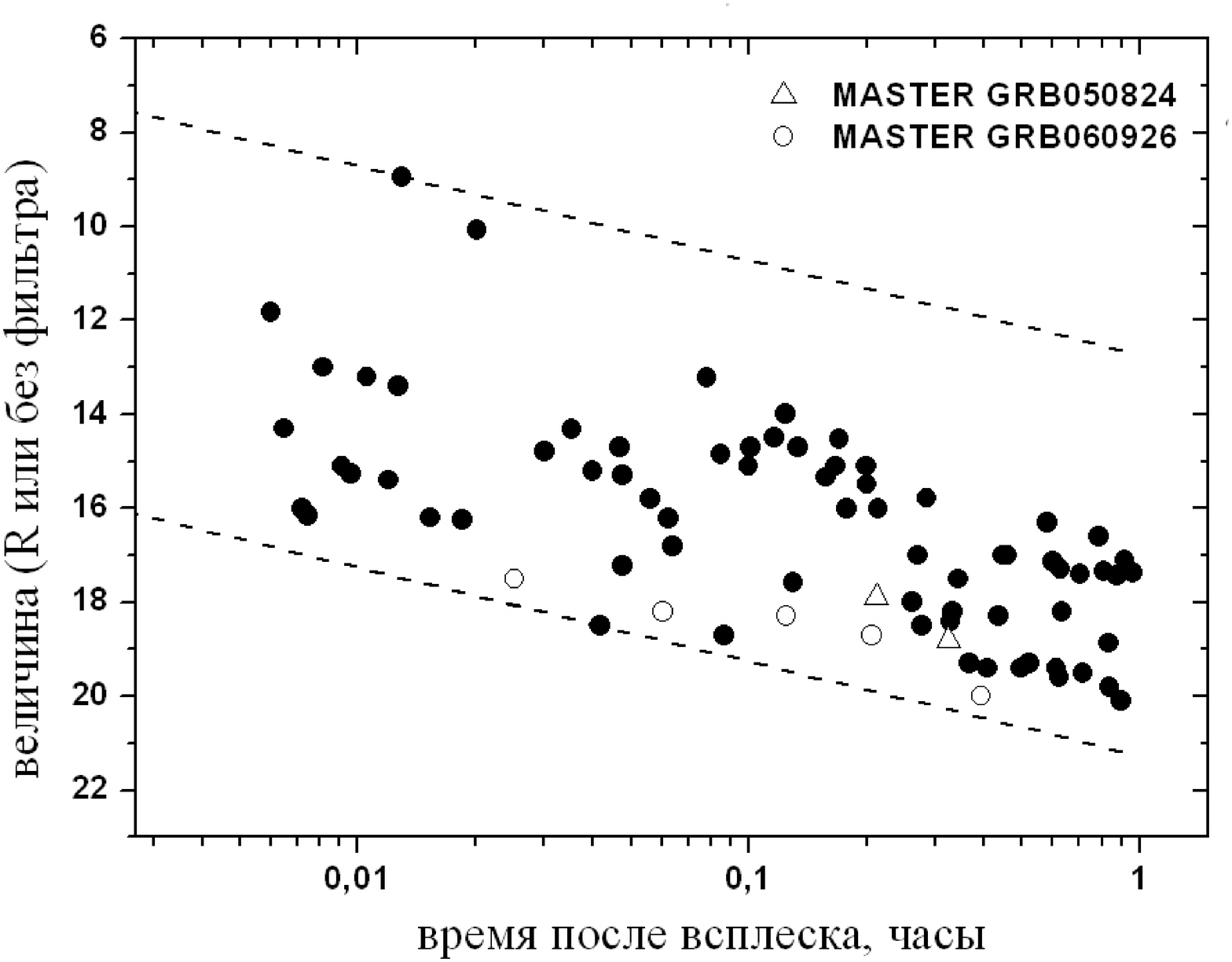,width=12cm}}
\caption{Master light curve for the GRB afterglows observed within an hour after the burst. The width of the band, shown by
dashed lines, corresponds to $\Delta m = 8\mag5$\label{fig:4}}
\end{figure}

\begin{figure}[ht]
\centerline{\psfig{figure=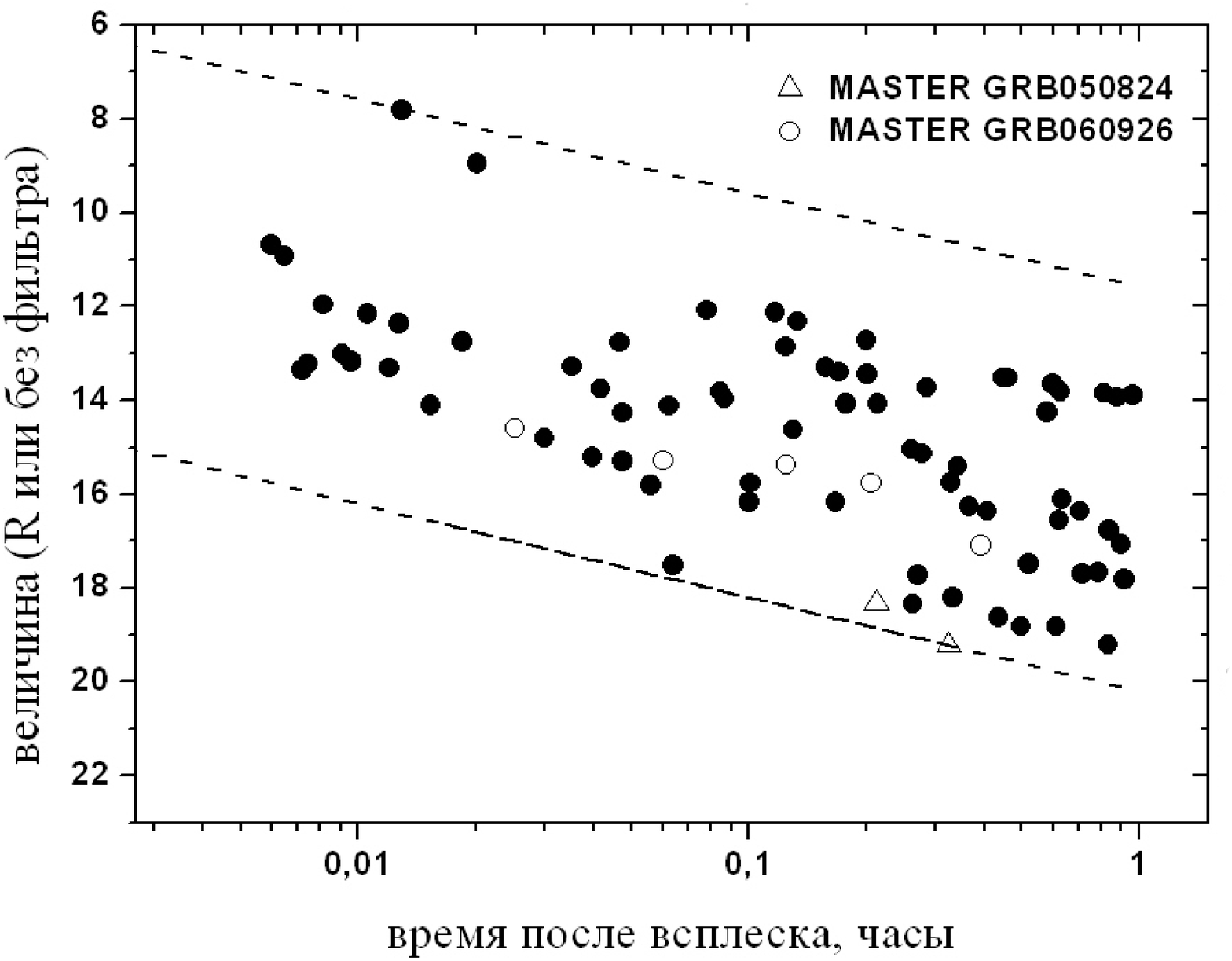,width=12cm}}
\caption{Synthetic optical light curve for GRBs normalized to the same redshift z (in white light or R). The width of the band is $\Delta m = 8\mag6$\label{fig:5}}
\end{figure}

\begin{figure}[ht]
\centerline{\psfig{figure=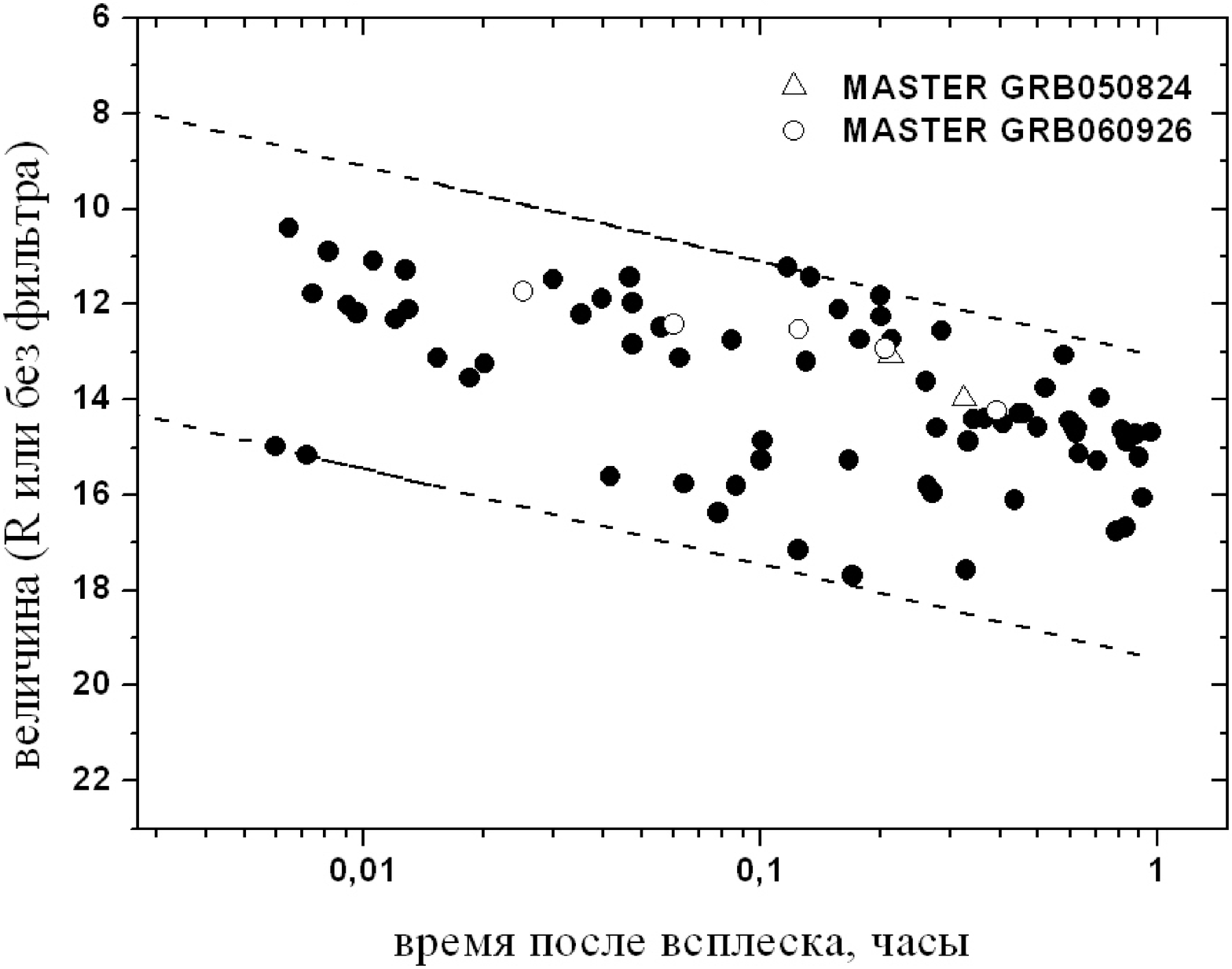,width=12cm}}
\caption{Synthetic light curve for GRBs in the optical and gamma-ray normalized to the same redshift $z$ and gamma-ray flux,
taking into account interstellar absorption. The width of the band has narrowed to $\Delta m = 6\mag3$\label{fig:6}}
\end{figure}

\begin{figure}[ht]
\centerline{\psfig{figure=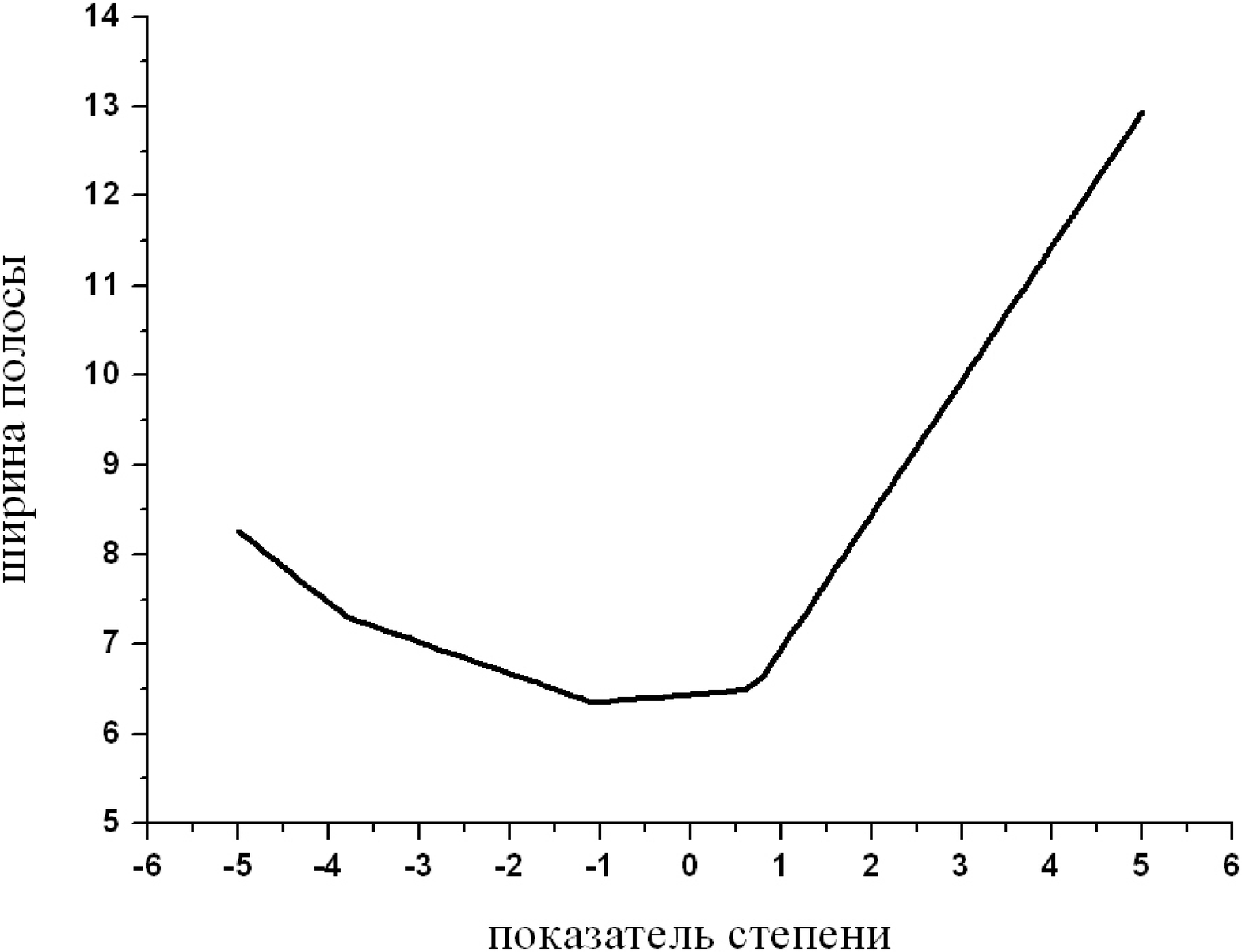,width=8cm}}
\caption{Dependence of the width of the band corresponding
to the synthetic GRB light curve on the power-law
index $\beta$ for the assumed spectrum. \label{fig:7}}
\end{figure}

\begin{figure}[h]
\centerline{\psfig{figure=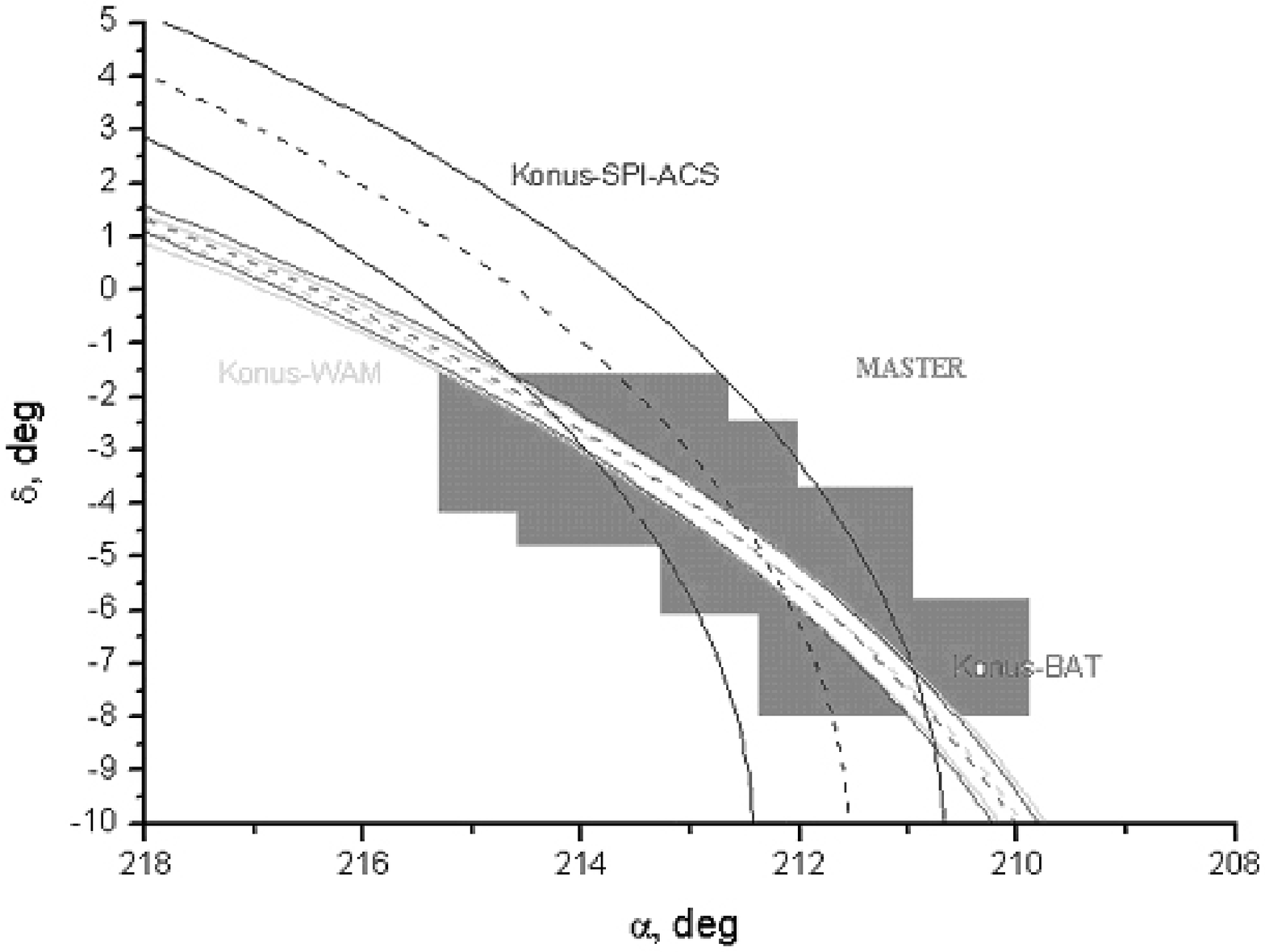,width=14cm}}
\caption{Preliminary large error box for GRB 060425 (kindly presented by the Konus-Wind group) together with areas
corresponding to theMASTER field of view (gray rectangles). \label{fig:8}}
\end{figure}

\begin{figure}[h]
\centerline{\psfig{figure=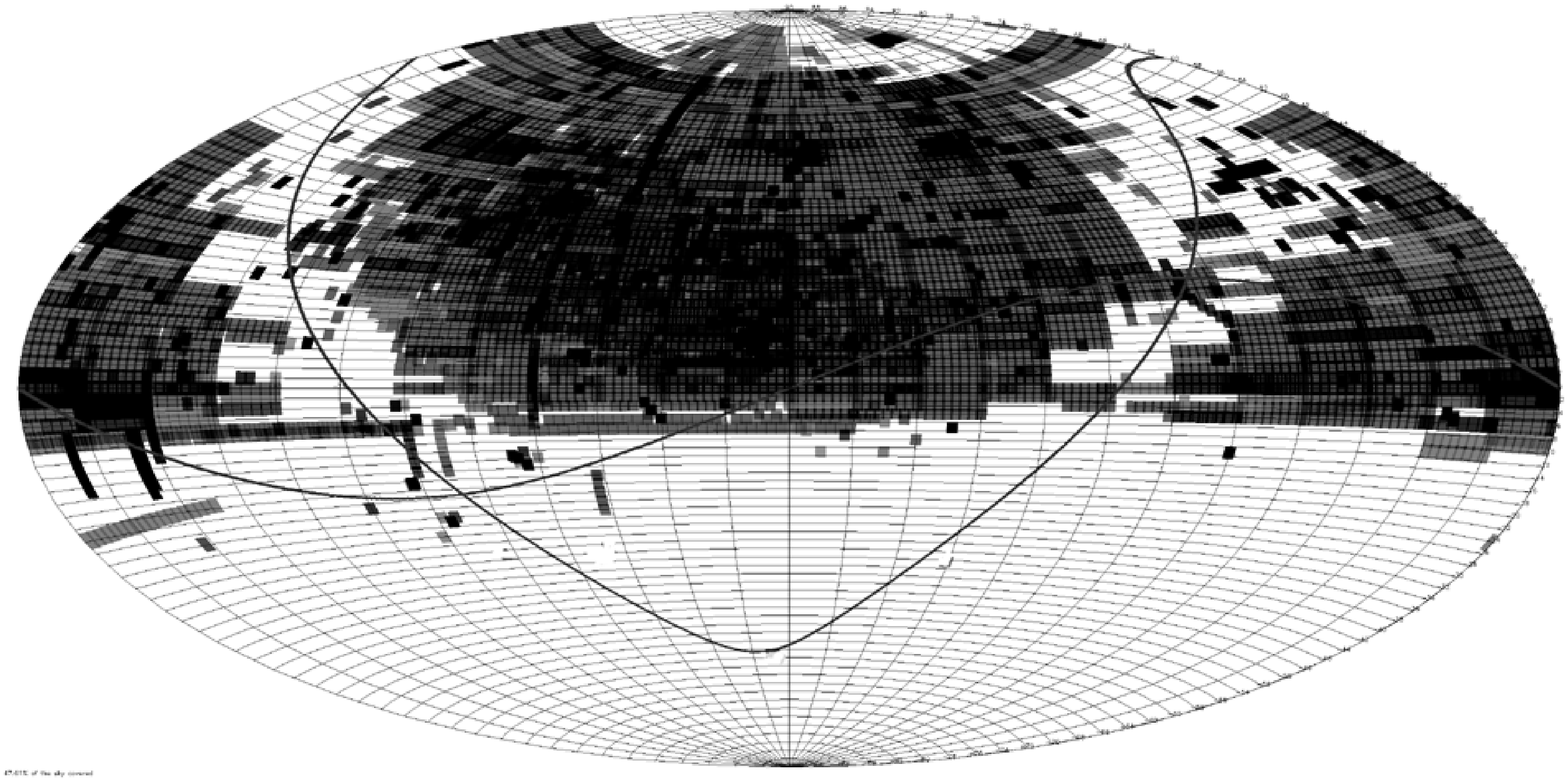,width=14cm}}
\caption{Sky coverage of survey images obtained on the mainMASTER telescope in October 2006. A dark square corresponds
to the size of a frame (six square degrees), and the darkness corresponds to the number of frames (no fewer than six). The lines
show the plane of the ecliptic and the Galactic plane. \label{fig:9}}
\end{figure}

\begin{figure}[h]
\centerline{\psfig{figure=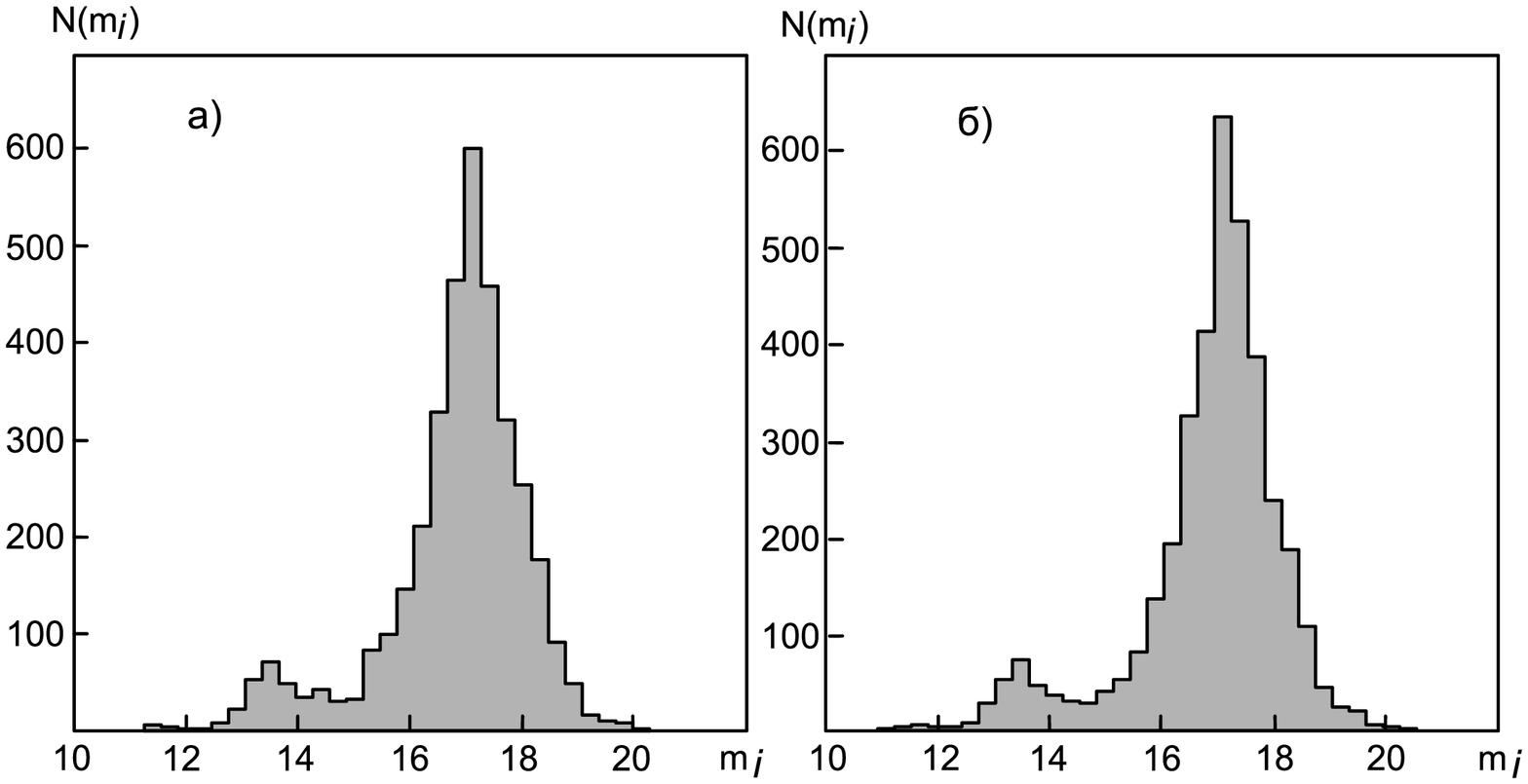,width=14cm}}
\caption{Distribution of limitingmagnitudes $m_i$  for the (a) first and (b) second pass of a sky survey by theMASTER telescope \label{fig:10}}
\end{figure}

\begin{figure}[h]
\centerline{\psfig{figure=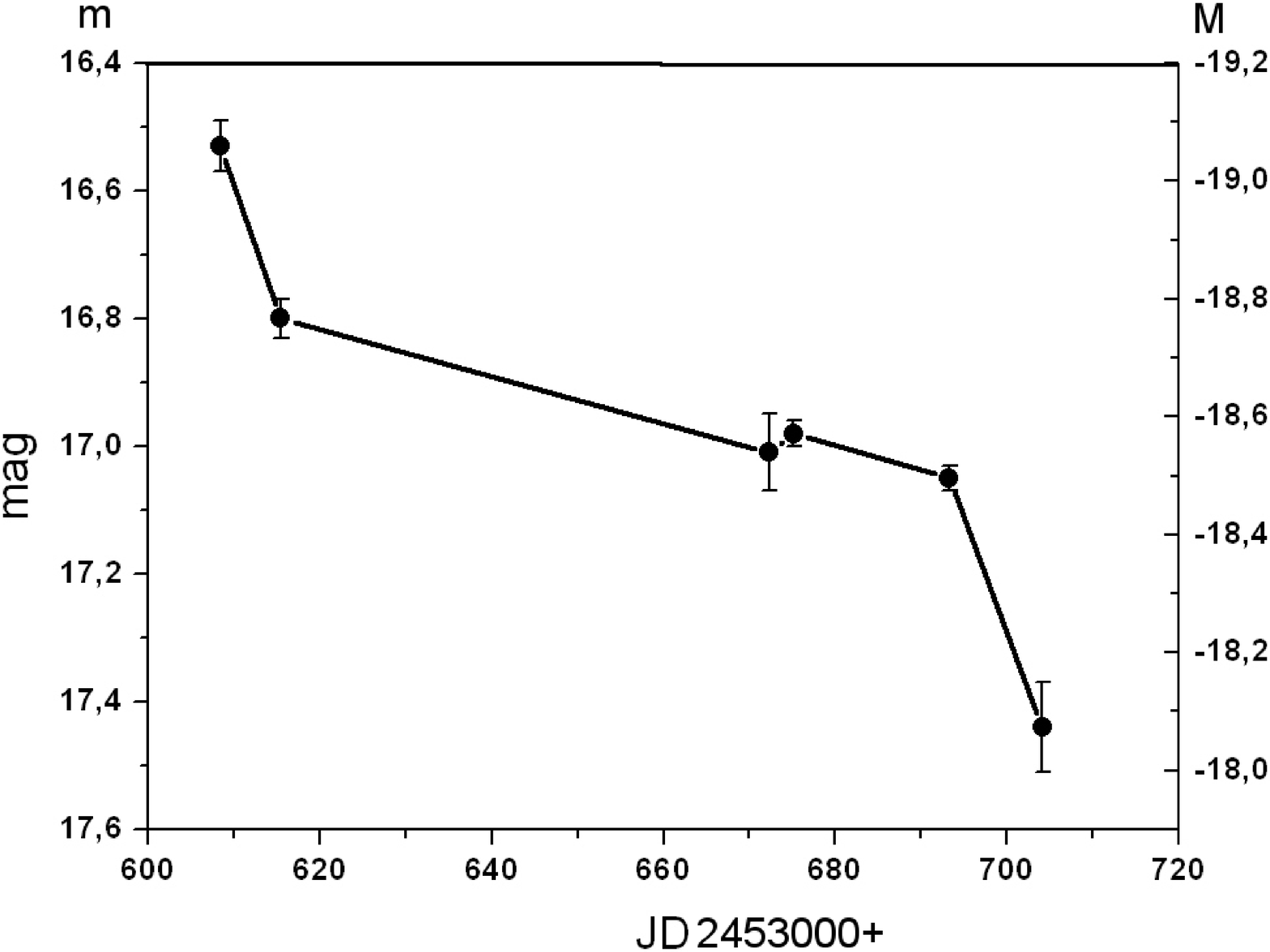,width=10cm}}
\caption{Decrease in brightness of SN 2005ee. The scale
to the left shows the instrumental magnitude, and the
scale to the right the absolute magnitude. \label{fig:11}}
\end{figure}

\begin{figure}[h]
\centerline{\psfig{figure=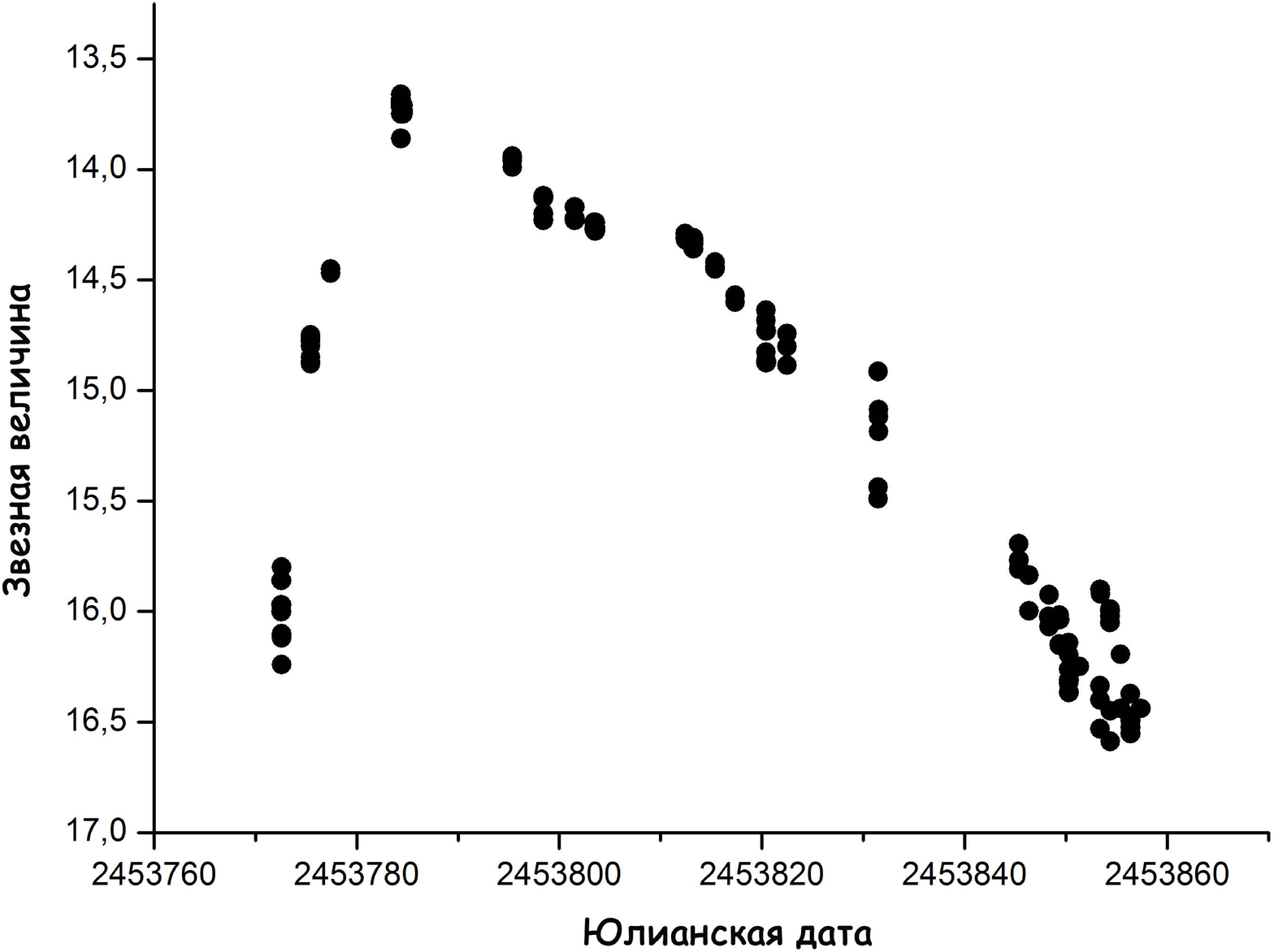,width=10cm}}
\caption{Light curve of SN 2006X \label{fig:12}}
\end{figure}

\end{document}